\documentstyle [epsfig,a4,12pt] {article}  
\def\chione{\tilde \chi_1^0}

\def\chionepm{\tilde \chi_1^\pm}

\def\chionep{\tilde \chi_1^+}

\def\m0{$m_{0}$}

\def\snm{{\tilde\nu_\mu}}

\def \susyq {supersymmetric }

\def\l {\lambda }

\def \ud {{1 \over 2} }

\def \bea {\begin{equation} }
\def \eea {\end{equation} }

\def \Eslash {E \kern-.9em\slash }
\def \pslash {p \kern-.5em\slash }
\def \kslash {k \kern-.5em\slash }
\newcommand{\rpv}{\mbox{$\not \hspace{-0.10cm} R_p$ }}

\newcommand{\LV}{\mbox{$\not \hspace{-0.15cm} L \hspace{0.1cm}$}}
\newcommand{\BV}{\mbox{$\not \hspace{-0.15cm} B \hspace{0.1cm}$}}

\def\dofig#1#2{\epsfxsize=#1\centerline{\epsfbox{#2}}}
\def\dofigs#1#2#3{\centerline{\epsfxsize=#1\epsfbox{#2}%
   \hfil\epsfxsize=#1\epsfbox{#3}}}

\def\bentarrow{\:\raisebox{1.3ex}{\rlap{$\vert$}}\!\rightarrow}

\def\dksl#1#2#3#4{
	\begin{equation}
	\begin{array}{r c l }
	#1 & \rightarrow & #2  \\
	 & & \bentarrow #3 \\
         & & \phantom{\bentarrow \;}  \bentarrow #4 
	\end{array}
        \label{eqq}
	\end{equation}
		}

\catcode`@=11 % This allows us to modify PLAIN macros.
\def \gsim{\mathrel{\mathpalette\@versim>}}
\def \lsim{\mathrel{\mathpalette\@versim<}}
\def \@versim#1#2{\lower0.4ex\vbox{\baselineskip\z@skip\lineskip\z@skip
     \lineskiplimit\z@\ialign{$\m@th#1\hfil##\hfil$%
     \crcr#2\crcr\sim\crcr}}}
\catcode`@=12 % at signs are no longer letters
\begin{document}           % End of preamble and beginning of text.
%================================================== TITLE =======
\begin{titlepage}
%\begin{tabbing}
%\` ATLAS Internal Note \\
%\` PHYS-No-XXX\\
%\` August 1999\\
%\end{tabbing}
%======================================================== TITLE =======
\begin{center}{\LARGE
\vspace{10 mm}
Resonant sneutrino production in
Supersymmetry with R-parity violation at the LHC 
\vspace{5mm}
}\end{center}\bigskip
%======================================================================
\vspace{5 mm}
\begin{center}
{\normalsize \bf
{G. Moreau\\}}
{\normalsize \em
Service de Physique Th\'eorique \\ 
CEA-Saclay, F91191, Gif-sur-Yvette, Cedex France\\
}
\vspace{2mm}
{\normalsize \bf
{E. Perez\\}}
{\normalsize \em
Service de Physique des Particules, DAPNIA \\
 CEA-Saclay, F91191, Gif-sur-Yvette, Cedex France\\
}
\vspace{2mm}
{\normalsize \bf
{G. Polesello\\}
}
{\normalsize \em
INFN, Sezione di Pavia, 
Via Bassi 6, Pavia, Italy\\
}
\end{center}
%======================================================================
\begin{abstract}
The resonant production of sneutrinos at the LHC via the 
R-parity violating couplings $\l ' _{ijk} L_i Q_j D^c_k$ 
is studied through its three-leptons signature.
A detailed particle level study of signal and background
is performed using a fast simulation of the ATLAS detector. 
Through the full reconstruction of
the cascade decay, a model-independent and precise measurement 
of the masses of the involved sparticles can be performed.
Besides, this  signature can be detected for a broad class of supersymmetric
models, and  for a wide range of values of several $\l ' _{ijk}$  
coupling constants. 
Within the MSSM, the production of
a 900~GeV sneutrino for $\lambda^{\prime}_{211}>0.05$, and of a 350~GeV sneutrino for
$\lambda^{\prime}_{211}>0.01$ can be observed within
the first three years of LHC running.\\ 

%================================================ end of abstract =====
\vspace{10 mm}
\parskip 3cm
\end{abstract}
%
%
%---------------
\end{titlepage}
%================================================ contents
\include{MEMO_defs}
%================================================ introduction
\newpage
%
%\hrule height 2pt      % draw a line
%\medskip
%---------------------------------------------- INTRODUCTION
%\newpage
%
%
%%%%%%%%%%%%%%%%%%%%%%%%%%%%%%%%%%%%%%%%%%%%%%% SECTION [********]
\section[intro]{Introduction}
%%%%%%%%%%%%%%%%%%%%%%%%%%%%%%%%%%%%%%%%%%%%%%%
The most general superpotential respecting the gauge 
symmetries of the Standard Model (SM) contains bilinear and trilinear terms
which are not taken into account in the 
Minimal Supersymmetric Standard Model (MSSM).
Restricting to the trilinear part, these additional terms read as~:
\begin{eqnarray}
W \supset \sum_{i,j,k} \bigg (\ud \l _{ijk} L_iL_j E^c_k+
\l ' _{ijk} L_i Q_j D^c_k+ \ud \l '' _{ijk} U_i^cD_j^cD_k^c   \bigg ), 
\label{super}
\end{eqnarray}
where $i,j,k$ are generation indices,
$L$ ($Q$) denote the left-handed leptons (quarks) superfields,
and $E^c$, $D^c$ and $U^c$ are right-handed superfields
for charged leptons, down and up-type quarks, respectively.

The first two terms in Eq.(\ref{super}) lead to violation of
the lepton number (\LV), while the last one implies violation
of the baryon number (\BV).
Since the simultaneous presence of \LV and \BV couplings
could lead to a too fast proton decay, a discrete multiplicative symmetry
which forbids the above terms in the superpotential
has been imposed by hand in the MSSM. 
This symmetry, called R-parity ($R_p$),
is defined as $R_p = (-1)^{3B + L + 2S}$, where
$B$, $L$ and $S$ respectively denote the baryon number,
the fermion number and the spin, such that $R_p=-1$ ($R_p = 1$)
for all supersymmetric (SM) particles.
However other solutions can ensure the proton stability,
e.g. if $L$ only is violated, or if only $U_i^cD_j^cD_k^c$ interactions
are allowed and the proton is lighter than
the Lightest Supersymmetric Particle (LSP).
Moreover, on the theoretical point of view, there is no clear preference,
e.g. in models inspired by Grand Unified or string theories,
between \rpv and $R_p$ conservation~\cite{Drein}.
It is thus mandatory to search for SUSY in both scenarios.

On the experimental side, the main consequence of \rpv
lies in the possibility for the LSP to decay into ordinary
matter.
% through a conversion to fermion-sfermion pairs
% followed by sfermion decay via a \rpv coupling.
This is in contrast to scenarios where $R_p$ is conserved,
in which the LSP is stable and escapes detection, leading to the
characteristic search for missing energy signals in direct 
collider searches.
Moreover, while in $R_p$ conserved models, 
the \susyq (SUSY) particles must be produced in pairs,
\rpv allows the single production of 
superpartners, thus enlarging the mass domain where
SUSY could be discovered.
In particular,
\rpv couplings offer the opportunity to resonantly
produce \susyq particles~\cite{Dim1,Dreinoss}.Although the \rpv
coupling constants are severely constrained by the low-energy 
experimental bounds \cite{Drein,Han,Bhatt,rapport,GDR,Alla}, the superpartner
resonant production can have significant cross-sections both at 
leptonic 
\cite{Han} and hadronic \cite{Dim2} colliders.
This is this possibility which is exploited throughout this paper. \\
The resonant production of \susyq particles is attractive for another reason:
Since its rate is proportional to a power $2$ of the relevant \rpv coupling, 
this reaction would allow an easier determination of the  
\rpv couplings than the pair production. 
In fact in the latter case,
the sensitivity on the \rpv coupling is
mainly provided by the displaced vertex analysis for the LSP decay,
which is difficult experimentally especially at hadronic colliders.

In this paper, we focus on the resonant SUSY particle production at 
the Large Hadron Collider (LHC) 
operating at a center of mass energy of $14$~TeV
with special reference to  the ATLAS detector.
At the LHC due to the continuous 
distribution of the centre of mass  energy of the
colliding partons, a parton-parton resonance can be probed over a wide  mass domain.
This is a distinct advantage over the situation at
lepton colliders, where the search for narrow resonances requires 
lengthy scans over the centre of mass energy of the machine.

At hadronic colliders, either a slepton or a squark can be produced at
the resonance through a $\l'$ or a $\l''$ coupling constant, respectively. 
In the hypothesis of a single dominant \rpv 
coupling constant, the resonant SUSY particle 
could decay through the same \rpv coupling as in the production, 
leading then to a two quarks final state for the hard process \cite{Bin,Dat,Oak,Rizz,Chiap}. 
In the case where both $\l'$ and $\l$ couplings 
are non-vanishing, the slepton produced via $\l'$ can 
decay through $\l$
giving rise to the same final state as in Drell-Yan process, 
namely two leptons  
\cite{Rizz,Kal,Kalp,SonPart}. However, for most of the values of the 
\rpv coupling constants allowed by present indirect searches, the decays  
of the resonant SUSY particle via gauge interactions are 
dominant if kinematically accessible \cite{Han}. 
In this favoured situation, typically, the
produced superpartner initiates a cascade decay ended by
the \rpv decay of the LSP. In case of a dominant
$\l''$ coupling constant, 
due to the \rpv decay of the LSP into quarks,
this cascade decay leads to multijet final states which have a
large QCD background \cite{Dim2,Bin}. Only  if  leptonic decays 
such as for instance $\tilde \chi^+_1 \to \bar l_i \nu_i \tilde \chi^0_1$ 
enter the cascade clearer signatures can be investigated \cite{Berg}. 
The situation is more favourable in the hypothesis of 
a single $\l'$ coupling constant, where the LSP can decay into a charged lepton,
% and multileptonic final states are often obtained.
allowing then multileptonic final states to be easily obtained.\\ 
We will thus assume a dominant $\l'_{ijk}$ coupling constant. 
At hadronic colliders, either a $\tilde \nu_i$ 
sneutrino or a $\tilde l_i$ charged slepton can be produced  
at the resonance via $\l'_{ijk}$ and the initial states are $d_j \bar d_k$
and $u_j \bar d_k$, respectively. 
The slepton produced at the resonance has two possible gauge 
decays, either into a chargino or a neutralino. 
In both cases particularly clean signatures can be observed. 
For example, the production of a neutralino together with a charged lepton 
resulting from the resonant charged slepton production can lead 
to the interesting like-sign dilepton topology \cite{Rich,Rich2}
since, due to its Majorana nature, the neutralino decays via
$\l'_{ijk}$ into a lepton as $\tilde \chi^0 \to l_i u_j \bar d_k$ 
and into an anti-lepton as $\tilde \chi^0 \to \bar l_i \bar u_j d_k$ 
with the same probability.

\begin{figure}
\begin{center}
\leavevmode
\centerline{\psfig{figure=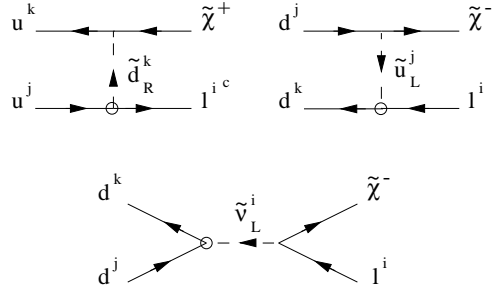,height=1.5in}}
\end{center}
\protect\caption{\em Feynman diagrams for the single chargino 
production at hadronic colliders via the $\l'_{ijk}$ coupling 
(symbolised by a circle in the figure). 
The arrows denote the flow of the particle momentum.}
\label{fig1}
\end{figure}

In this article, we consider the single lightest chargino
production at LHC as induced by the resonant sneutrino production 
$ p p \to \tilde \nu_i \to \tilde \chi^+_1 l_i$. The single 
$\tilde \chi^{\pm}_1$ production also  receives contributions from the
$t$ and $u$ channel squark exchange diagrams shown in Figure~\ref{fig1}.
In many models, the $\tilde \chi^0_1$ neutralino is the LSP 
for most of the SUSY parameter space. 
In the hypothesis of a $\tilde \chi^0_1$ LSP, 
the produced $\tilde \chi^{\pm}_1$ chargino mainly decays into 
the neutralino as $\tilde \chi^{\pm}_1 \to \tilde \chi^0_1 q_p \bar q'_p$
or as $\tilde \chi^{\pm}_1 \to \tilde \chi^0_1 l^{\pm}_p \nu_p$. 
The neutralino then decays via $\l'_{ijk}$ as 
$\tilde \chi^0_1 \to l_i u_j \bar d_k, \ \bar l_i \bar u_j d_k$ or as 
$\tilde \chi^0_1 \to \nu_i d_j \bar d_k, \ \bar \nu_i \bar d_j d_k$. 
We concentrate on the decays of both the chargino and the neutralino 
into charged leptons, which lead to a three leptons final state.
This signature has a low Standard Model background, and 
allows the reconstruction of the whole decay chain, thus providing
a measurement of some parameters of the SUSY model.

%------------   FIGURE 1 : Feynman diagrams ----------------------
%------------------------------------------------------------------

%%%%%%%%%%%%%%%%%%%%%%%%%%%%%%%%%%%%%%%%%%%%%%%%%%%%%%%%%%%%%%%%%%%%%%
% \include{signal}
%%%%%%%%%%%%%%%%%%%%%%%%%%%%%%%%%%%%%%%%%%%%%%%%%%%%%%%%%%%%%%%%%%%%%%

%%%%%%%%%%%%%%%%%%%%%%%%%%%%%%%%%%%%%%%%%%%%%%%%%%%%%%%%%%%%%%%%%%%%%%
\section{The signal}
\label{secmssm}
%%%%%%%%%%%%%%%%%%%%%%%%%%%%%%%%%%%%%%%%%%%%%%%%%%%%%%%%%%%%%%%%%%%%%%

%==================================
\subsection{Theoretical framework}
%==================================

Our theoretical framework in sections \ref{secmssm} and \ref{analysis}
will be the \rpv extension of the Minimal Supersymmetric Standard Model.
In Section \ref{reach} we will also give results in the Minimal 
Supergravity (mSUGRA) model. The MSSM parameters are the following.
$M_1$, $M_2$ and $M_3$ are the soft-SUSY breaking mass terms
for the bino,  the wino and the gluino, respectively.
$\mu$ is the Higgs mass parameter.
$\tan \beta=<H_u>/<H_d>$ is the ratio of the vacuum expectation values (vev)
for the two-Higgs doublet fields. $A_t$, $A_b$ and $A_{\tau}$ are the
third generation soft-SUSY breaking trilinear couplings. In fact, since
these
trilinear couplings are proportional to the fermion masses one
can neglect the first two generations couplings without any phenomenological
consequence in this context. Finally, $m_{\tilde q}$, $m_{\tilde l}$
and $m_{\tilde \nu}$ are the squark, slepton
and sneutrino mass, respectively.
The value of the squark mass enters our study mainly
in the determination
of the relative branching ratios of the $\tilde \chi^0$ into lepton or
neutrino and of the $\tilde \chi^{\pm}$
into $\tilde \chi^0$ + quarks or $\tilde \chi^0$ + leptons. The remaining
three
parameters $m_{H_u}^2$, $m_{H_d}^2$ and the
soft-SUSY breaking bilinear coupling $B$ are
determined through the electroweak symmetry breaking conditions
which are two necessary minimisation conditions of the Higgs potential.

We choose to study the case of a single dominant
$\lambda_{2jk}^{\prime}$ allowing the reactions
$p p \rightarrow {\tilde{\chi}}^{\pm} \mu^{\mp}$.
In section~\ref{analysis} 
the analysis will be performed explicitly for the
$\lambda_{211}^{\prime}$ coupling, since it corresponds to
the hard subprocess $d \bar d \to \tilde \chi^{\pm}_1 \mu^{\mp}$
which offers the highest partonic luminosity. We will 
take $\lambda_{211}^{\prime}$=0.09, the upper value
allowed by indirect bound: 
$\lambda_{211}^{\prime}<0.09(m_{\tilde d_R}/100GeV)$  \cite{Drein}
for a squark mass of 100~GeV. A 
quantitative discussion will be given below for the general case
of a single dominant $\lambda_{2jk}^{\prime}$ coupling constant.
We will not treat explicitly the $\lambda^\prime_{1jk}$ couplings
which are associated to the $\tilde \chi^{\pm}$-$e^{\mp}$ production, 
since the low-energy bounds on these couplings 
are rather more stringent than the constraints
on $\lambda^\prime_{2jk}$ and $\lambda^\prime_{3jk}$ \cite{Drein}.
However, the three-leptons analysis from sneutrino production
should give similar sensitivities on the $\lambda^\prime_{1jk}$ and
$\lambda^\prime_{2jk}$
couplings since isolation cuts will be included in the selection criteria
for the leptons.
We will not perform the
analysis of the $\lambda^\prime_{3jk}$ couplings which
correspond to the $\tilde \chi^{\pm}$-$\tau^{\mp}$ production.
A technique for mass reconstruction in the ATLAS detector
using the hadronic decays of the $\tau$
has been demonstrated in \cite{ianfrank}. The detailed
experimental analysis needed to extract a signal is
beyond the scope of this work.
Besides, in this case the sneutrino and chargino mass reconstruction
studied in Section \ref{secana} is spoiled by the neutrinos produced
in the $\tau$ decay.

%=====================================================
\subsection{Single chargino production cross-section}
%=====================================================
\label{xsec}

In order to establish the set of models in which
the analysis presented below can be performed, 
we need to study the variations of the single chargino production
rate $\sigma(p p \to \tilde \chi^{\pm} \mu^{\mp})$ with the
MSSM parameters.

%-------------   FIGURE 2 : sigma(chip mu) versus mu --------------------
\begin{figure}
\begin{center}
\dofig{15cm}{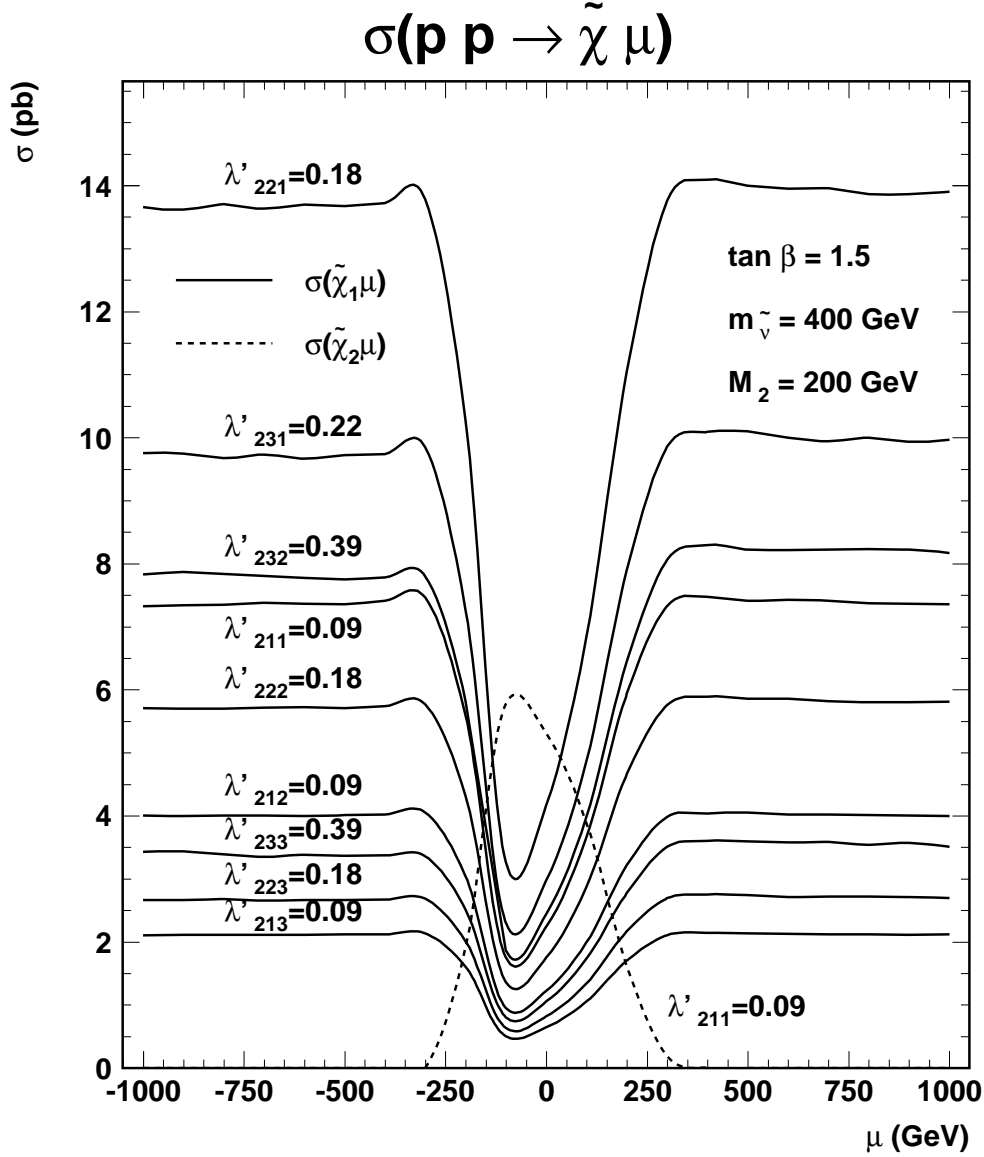}
\protect\caption{\em
Cross-sections for the $\tilde \chi_1^{\pm}$-$\mu^{\mp}$
production as a function of the $\mu$ parameter
through various $\lambda'_{2jk}$ couplings,
for $\tan\beta$=1.5, $M_2=200$~GeV and
$m_{\tilde \nu}=400$~GeV. In the case of
$\lambda'_{211}$ the cross-section for $\tilde \chi_2^{\pm}$-$\mu^{\mp}$
is also shown as the dashed line. The values of the
\rpv couplings have been chosen equal to:
$\lambda^\prime_{211}=0.09$, $\lambda^\prime_{212}=0.09$,
$\lambda^\prime_{213}=0.09$, $\lambda^\prime_{221}=0.18$,
$\lambda^\prime_{222}=0.18$, $\lambda^\prime_{223}=0.18$,
$\lambda^\prime_{231}=0.22$, $\lambda^\prime_{232}=0.39$ 
and $\lambda^\prime_{233}=0.39$, which
correspond to the low-energy limits
for a sfermion mass of $100$~GeV \cite{Drein}.}
\label{mssmu}
\end{center}
\end{figure}
%--------------------------------------------------------------------------

In Figure~\ref{mssmu}, we present the cross-sections
for the $\tilde \chi_1^{\pm}$-$\mu^{\mp}$ production
through several $\lambda^\prime_{2jk}$ couplings
as a function of the $\mu$ parameter
for the fixed values: $\tan\beta$=1.5, $M_2=200$~GeV, and
$m_{\tilde \nu}=400$~GeV. For this choice of parameters
and independently of $\mu$,
the chargino ${\tilde{\chi}}^{\pm}_{1}$ is 
lighter than the $\tilde{\nu}$. In this case the contributions of
squark exchange in the $t$ and $u$ channels
are negligible compared to the resonant process
so that the $\tilde \chi_1^{\pm}$-$\mu^{\mp}$ production cross-section
does not depend on the squark mass.
The values for the considered \rpv coupling constants have been 
conservatively taken equal
to the low-energy limits for a sfermion mass of $100$~GeV \cite{Drein}.
The cross-sections scale as $\lambda_{2jk}^{\prime 2}$. \\
We see on this Figure that the dependence of the rates on $\mu$ is smooth
for $| \mu | > M_2$.
This is due to the weak dependence of the $\tilde \chi_1^{\pm}$ 
mass on $\mu$ in this domain.
In contrast, we observe a strong decrease of the rate in the region 
$| \mu | < M_2$
where the
$\tilde \chi_1^{\pm}$ chargino is mainly composed by the higgsino.
Most of the small $ |\mu |$ domain  ($\vert \mu \vert$ smaller than
$\sim 100$~GeV for $\tan \beta=1.41$ and $m_0=500$~GeV)
is however excluded by the present LEP limits~\cite{aleph2}. \\
We also show as a dashed line on the plot
the rate for the $\tilde \chi_2^{\pm}$-$\mu^{\mp}$ production through the
$\lambda^\prime_{211}$ coupling.
The decrease of the $\tilde \chi_2^{\pm}$ production rate with increasing
$\vert \mu \vert$
is due to an increase of the $\tilde \chi_2^{\pm}$ mass.
We will not consider the contribution to the three-leptons final state from
the
$\tilde \chi_2^{\pm}$ production since the rate 
becomes important only for a very limited 
range of small $\vert \mu \vert$ values not yet excluded by LEP data. \par
Figure~\ref{mssmu} also allows to compare the sensitivities that
can be reached on various $\lambda'_{2jk}$ couplings
using the single chargino production.
If we  compare for instance the cross-sections of the
$\tilde \chi_1^{\pm}$ production
via $\lambda^\prime_{211}$ and $\lambda^\prime_{221}$
at $\mu=-500$~GeV,
we can see that for equal values of the \rpv
couplings the ratios between the cross-sections
associated to $\lambda^\prime_{211}$ and
$\lambda^\prime_{221}$
is $\sim 2.17$. Therefore, the sensitivity that can be obtained on
$\lambda^\prime_{221}$ is only $\sim \sqrt {2.17}$ times
weaker than the sensitivity on $\lambda^\prime_{211}$,
for a $400$~GeV sneutrino.
Note that the cross-section ratio, and hence the
scaling to be applied, in order to infer
from the reach on $\lambda'_{211}$ the sensitivity on
another coupling $\lambda'_{2jk}$, depends on the 
sneutrino mass.
%Indeed, since the single
%chargino production via $\lambda'_{2jk}$ mainly proceeds, as already mentioned,
%through the resonant sneutrino production 
%$d_k \bar d_j \to \tilde \nu_{\mu}$,
%some of the ratios between the different cross-sections presented in
%Figure~\ref{mssmu}
%would be modified by changing the sneutrino mass.
The reason is that the evolution of the parton densities
with the $x$-Bjorken variable is different for sea quark and valence quark
and for different quark flavours.

%----------   FIGURE 3 : sigma(chip mu) in (m_chip, m_snu) plane ------
\begin{figure}[htb]
\begin{center}
\dofig{9cm}{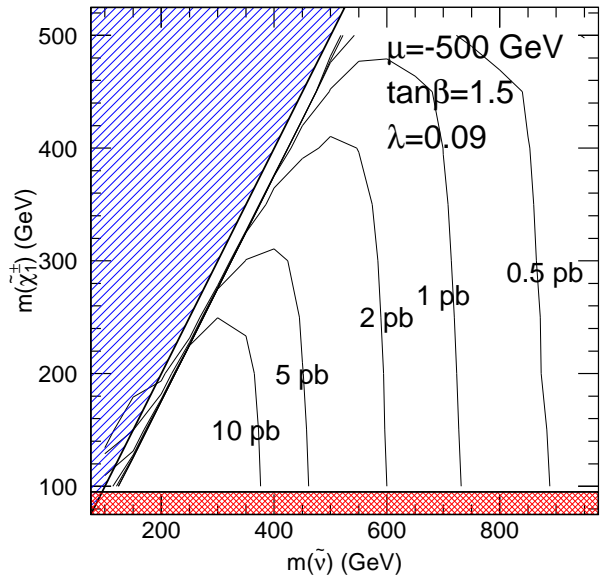}
\protect\caption{\em Cross-section for $\chionepm$-$\mu^{\mp}$
production as a function of $m_{\tilde\nu}$ and $m_{\chionepm}$
in the MSSM for the choice of values
$\mu=-500$~GeV, $\tan\beta=1.5$ and
$\lambda^\prime_{211}=0.09$.
The hatched region at the upper
left corresponds to $m_{\tilde \nu}<m_{\tilde\chi^{\pm}_1}$.
The cross-hatched region at low $m_{\chionepm}$ is excluded by the
preliminary LEP results at $\sqrt{s}=196$~GeV \cite{aleph2}.}
\label{mssmxse}
\end{center}
\end{figure}
%----------------------------------------------------------------------

In order to study the dependence of the cross-section on
the masses of the involved sparticles,
the parameters $m_{\tilde \nu}$ and $M_2$ were varied,
and the other model parameters
affecting the cross-section were fixed at the values:
$\lambda^\prime_{211}=0.09$,
$\mu=-500$~GeV and $\tan \beta=1.5$. 
The cross-section for $\chionepm$-$\mu^{\mp}$ production
as a function of
$m_{\tilde\nu}$ and $m_{\chionepm}$ is shown in Figure~\ref{mssmxse}.
Since the $\chionepm$ mass is approximately
equal to $M_2$ as long as $M_2< \vert \mu \vert$, and becomes equal 
to $\vert \mu \vert $ for $M_2 > \vert \mu \vert$,
we studied $\tilde\nu$ masses between 100 and 950 GeV, and values
of
$M_2$ between 100 and 500~GeV. 
For increasing $m_{\tilde\nu}$ the cross-section decreases
due to a reduction of the partonic luminosity.
A decrease of the cross-section is also observed for 
$m_{\chionepm}$  approaching $m_{\tilde\nu}$,
since the phase space factor of the decay $\tilde\nu\to\chionepm\mu^{\mp}$
following the resonant sneutrino production
is then suppressed. 
In the region $m_{\chionepm}>m_{\tilde\nu}$,
the chargino production still receives contributions from 
the $s$ channel exchange of a virtual sneutrino, as well as
from the $t$ and $u$ channels squark exchange 
which in that case also contribute significantly.
However, in this phase space domain where the resonant sneutrino 
production is not accessible, the cross-section is considerably 
reduced.

Finally, the single chargino production rate depends weakly on the
$A$ trilinear couplings. 
Indeed, only the $t$ and $u$ channels squark exchange,
varying with the squark mass which can be influenced by $A$,
depends on these couplings.
The dependence of the rate on the $\tan \beta$ parameter
is also weak.

%==========================================
\subsection{Three leptons branching ratio}
%==========================================

We calculate the total three leptons rate by multiplying
the single chargino cross-section by the 
chargino branching ratio, since we neglect the width of the chargino.
The three-leptons  final state is generated by the cascade decay
$\tilde \chi^{\pm}_1 \to \tilde \chi^0_1 l^{\pm}_p \nu_p$,
$\tilde \chi^0_1 \to \mu u d$.
For $m_{\tilde \nu},m_{\tilde l},m_{\tilde q},
m_{\tilde \chi^0_2}>m_{\tilde \chi^{\pm}_1}$,
the chargino decays mainly 
into a real or virtual $W$ and a $\chione$ and hence 
its branching fraction for the decay into leptons (lepton=$e,\mu$)
is $\sim 22 \%$.
%The branching ratio for the leptonic decay 
%of the chargino is typically around $30\%$ 
%and is smaller than for the hadronic decay
%$\tilde \chi^{\pm}_1 \to \tilde \chi^0_1 \bar q_p q'_p$
%because of the colour factor.
In particular kinematic configurations, the \rpv
modes can compete with the gauge couplings, affecting the
$\chionepm$ branching fractions. However, this does not happen
as long as the chargino is sufficiently heavier than the neutralino,
as is the case for example 
in supergravity inspired models.
When $\tilde \chi^0_1$ is the LSP, 
the branching ratio $B(\tilde \chi^0_1 \to \mu u d)$ ranges between
$\sim 40\%$ and $\sim 70\%$.
For values of $\vert \mu \vert$ much smaller than $M_2$
the other allowed decay $\tilde \chi^0_1 \to \nu_{\mu} d d$  becomes dominant,
spoiling the three-leptons signature.

%%%%%%%%%%%%%%%%%%%%%%%%%%%%%%%%%%%%%%%%%%%%%%
\section{Experimental analysis}
\label{analysis}
%%%%%%%%%%%%%%%%%%%%%%%%%%%%%%%%%%%%%%%%%%%%%%%
%
%================================
\subsection{Mass reconstruction}
\label{secana}
%================================
%
The analysis strategy is based on the exploitation of the decay chain:
\dksl{\snm}{\chionep \;\mu^-}{\chione\;\;W^+\rightarrow e^+(\mu^+)\nu}{\mu^{\pm} \; q\; \bar q^{\prime}}
which presents a sequence of three decays which can be fully reconstructed. 
The strong kinematic constraint provided by the masses of the three sparticles in 
the cascade is sufficient to reduce the contribution of the different
background sources well below the signal rate.

The signal events were generated with a version of the
SUSYGEN MonteCarlo~\cite{susygen} modified to allow the generation
of $pp$ processes. The hard-subprocess
$q \bar{q'} \rightarrow \tilde{\chi}^{\pm} \mu^{\mp}$ is first
generated according to the full lowest order matrix elements
corresponding to the diagrams depicted in Figure~\ref{fig1}.
Cascade decays of the $\tilde{\chi}$'s are performed according
to the relevant matrix elements. The parton showers
approach~\cite{JETSET74} relying on the DGLAP~\cite{DGLAP}
evolution equations is used to simulate QCD radiations
in the initial and final states, and 
the non-perturbative part of the hadronization
is modeled using string fragmentation~\cite{JETSET74}. The 
events were then processed through the program ATLFAST \cite{ATLFAST}, 
a parameterized simulation of the ATLAS detector response.

In this section, the analysis will be performed for the
\rpv coupling $\l'_{211}=0.09$ and for the following MSSM point:\\
$M_1=75$ GeV, $M_2=150$ GeV, $\mu=-200$ GeV, $\tan \beta=1.5$, 
$A_t=A_b=A_{\tau}=0$, $m_{\tilde f}=300$ GeV. \\
For this set of MSSM parameters, the masses of the relevant gauginos are:
%m_{\chitwo}=161.2~{\mathrm GeV}
%m_{\tilde \chi_2^{\pm}}=221.4~{\mathrm GeV}
$$
m_{\chione}=79.9~{\mathrm GeV} \;\; m_{\tilde \chi_1^{\pm}}=162.3~{\mathrm GeV}
$$
and the $\tilde \chi_1^{\pm}$ decay into an on shell $W$ has
a branching ratio of order $100 \%$.
The total cross-section for the resonant sneutrino 
production $pp \to \tilde \nu$ is 37~pb. If we include 
the branching fractions into the three leptons,  the cross-section
is 3.3~pb, corresponding to $\sim 100000$ events for the standard integrated
luminosity of 30~fb$^{-1}$ for the first three years of LHC data taking.

The signal is  characterised by the presence of three isolated leptons and two jets.
For the initial sample selection we require that:
\begin{itemize}
\item
Exactly three isolated leptons are found in the event, with 
$p_T^1>20$~GeV, $p_T^{2,3}>10$~GeV, where $p_T$ is the momentum
component in the plane perpendicular to the beam direction,
and pseudorapidity  $|\eta|<2.5$.
\item
At least two of the three leptons must be muons.
\item
At least two jets with $p_T>15$~GeV are found. 
\item
The invariant mass of any $\mu^+\mu^-$ pair is 
outside $\pm6.5$~GeV of the $Z$ mass. 
\end{itemize}
The isolation prescription on the leptons is necessary to reduce the 
background from the semileptonic decays of heavy quarks, 
and consists in requiring an energy deposition of less than 10~GeV not
associated with the lepton in a pseudorapidity-azimuth ($\eta-\phi$) cone 
of opening $\Delta R=0.2$ around the lepton direction.\par
The efficiency for  these cuts, after the branching fractions have been taken 
into account,  is $\sim 25\%$, where half of the loss comes from requiring
three isolated leptons, 
and the other half is the loss of jets from 
$\chione$ decay either because they are not reconstructed, or
because the two jets from the decay are reconstructed 
as a single jet . The $Z$ mass cut gives a 10$\%$ loss in statistics. 
In order to avoid the combinatorial background from additional 
QCD events we further require that no third jet with $p_T>15$~GeV
is reconstructed in the event. The efficiency after this cut is $\sim15\%$.\par
The reconstruction of the sparticle masses could be performed either
starting from the $\chione$ reconstruction and going up the decay chain,
or trying to exploit the three mass constraints at the same time.
We choose the first approach which is not optimal, but allows a clearer
insight into the kinematics of the events.
%
%----------  FIGURE 4 : mu-j-j mass for evts config 1 -------------------
\begin{figure}
\begin{center}
\dofigs{0.5\textwidth}{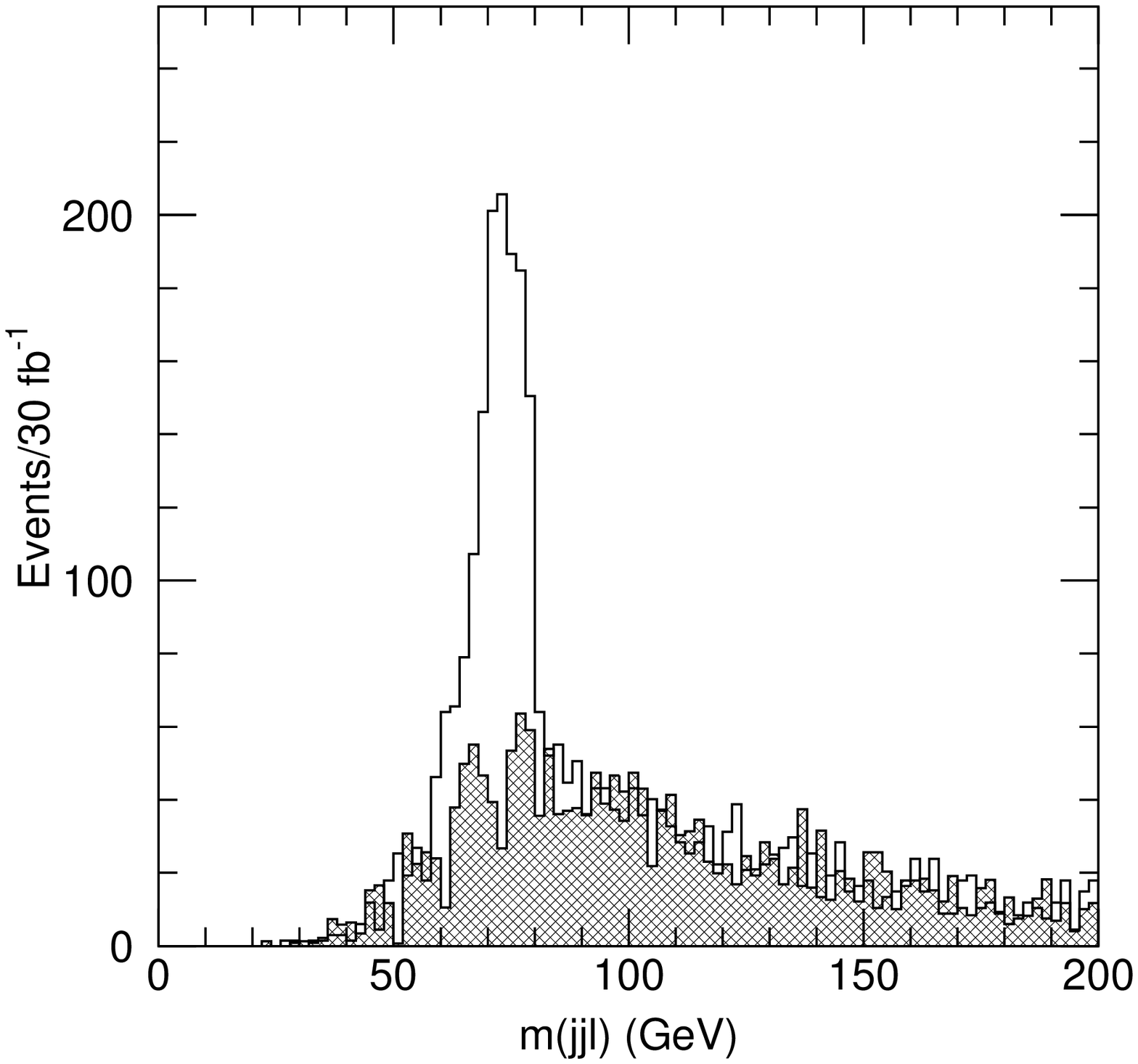}{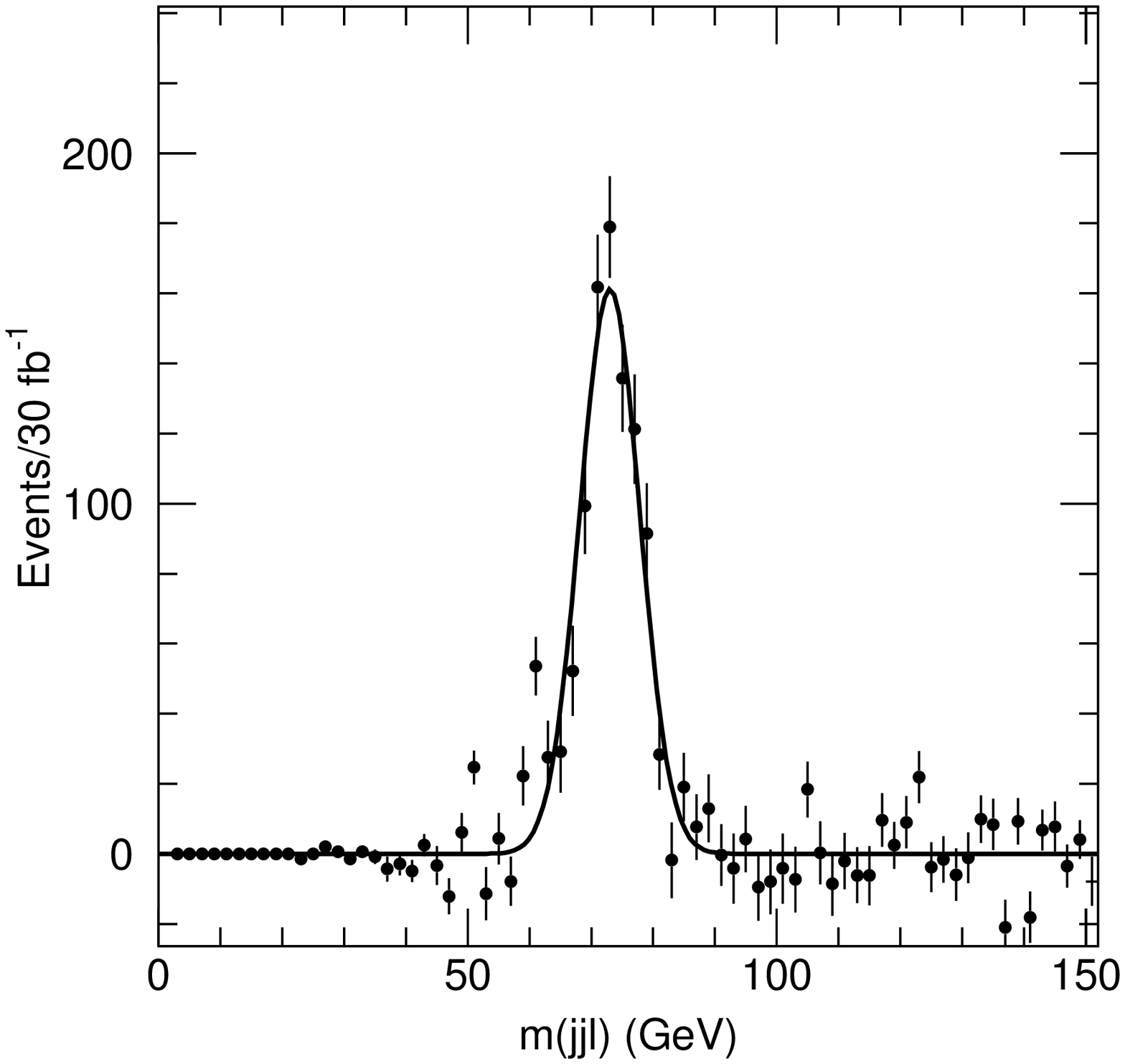}
\protect\caption{
\em $\mu$-jet-jet invariant mass for events in configuration 1. (see text)
Left: exclusive two jet events with superimposed (hatched) 
the combinatorial background 
Right: $\chione$ peak after background
subtraction.}
\label{figchi01}
\end{center}
\end{figure}
%-------------------------------------------------------------------------

The first step in reconstruction of the $\chione \rightarrow \mu$~jet~jet 
is the choice of the correct muon to attempt the reconstruction. 
The three leptons come in the following  flavour-sign configurations (+ charge conjugates):
\begin{enumerate}
\item
$\mu^- e^+\mu^+$
\item
$\mu^- e^+\mu^-$
\item
$\mu^-\mu^+\mu^+$
\item
$\mu^-\mu^+\mu^-$
\end{enumerate}
where the first lepton comes from the $\snm$, the second one from 
the $W$, and the third one from the $\chione$ decay, corresponding to
three final state signatures : 1)  two opposite sign muons and an electron
\footnote{Here and in the following, ``electron'' stands for both $e^+$ and $e^-$.}, 
2) two same-sign muons and an electron, 3-4) three muons. The configuration 
with three same-sign muons does not correspond to the required signature
and is rejected in the analysis.
For signature 1) 
the muon produced in the $\chione$ decay is defined as the one which has 
the same sign as the electron. For configuration 2) both muons must be tested
to reconstruct the $\chione$. For configuration 3-4), the $\chione$ muon must be
one of the two same-sign ones.\par
%
%-----------  FIGURE 5 : lepton-j-j mass, for 5 values of M_chi0
\begin{figure}
\begin{center}
\dofig{16cm}{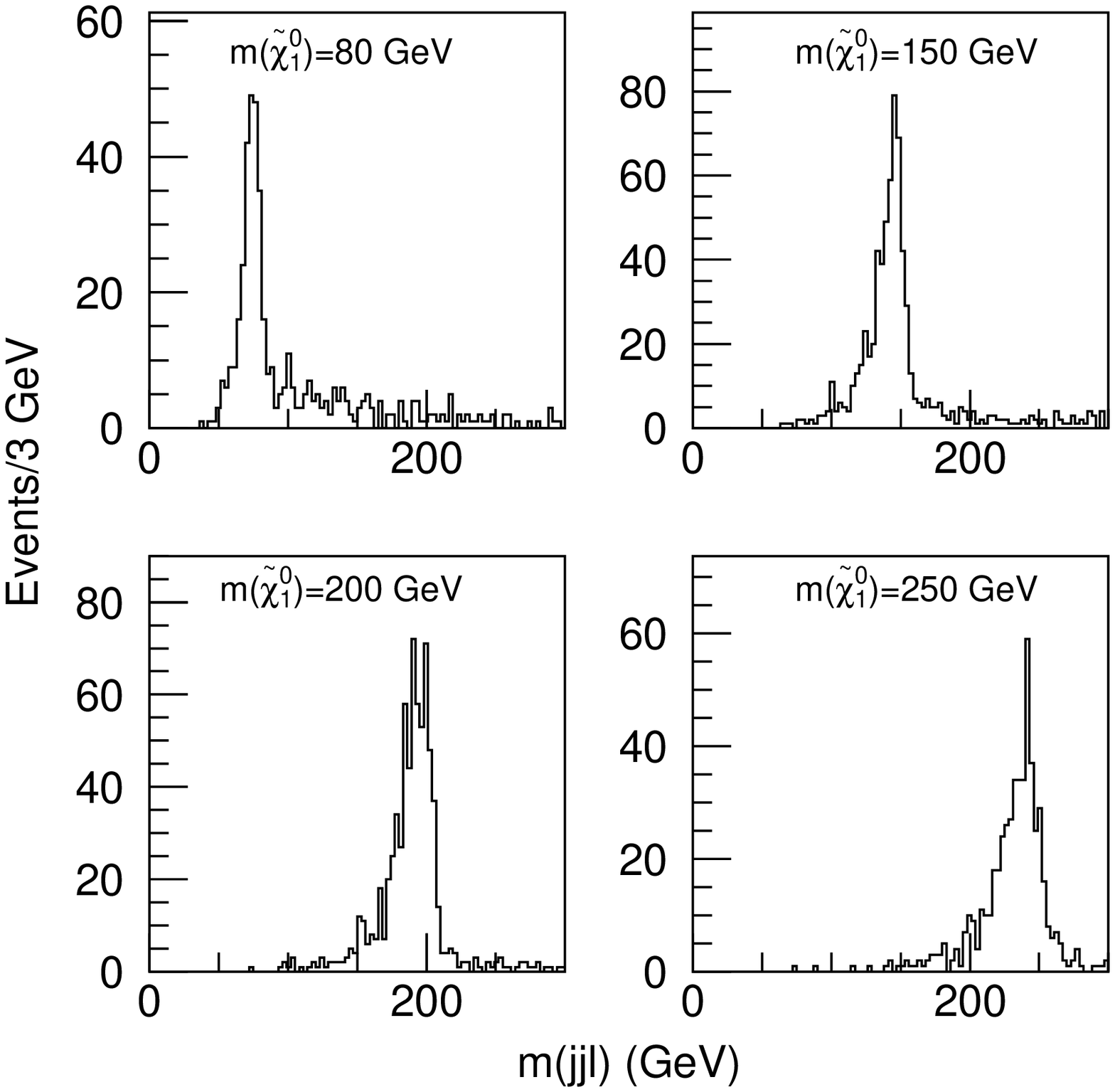}
\protect\caption{\em Lepton-jet-jet invariant mass for exclusive two-jet 
events where the $\chione$ lepton is
uniquely defined, for four different values of the $\chione$ mass:
$m_{\chione}$=80, 150, 200 and 250~GeV.
In all cases the sneutrino mass is set at 500~GeV, \mbox{$-\mu=M_2=2M_1$}, 
and \mbox{$\tan\beta=1.5$},
yielding a $\chionepm$ mass twice the $\chione$ mass. 
All the sfermion masses are set to 500~GeV.\
The normalisation is arbitrary.}
\label{figchi}
\end{center}
\end{figure}
%-------------------------------------------------------------------------

In order to minimise the combinatorial background 
we start the reconstruction from signature
1) where each lepton is unambiguously attributed to a step in the decay.
The distribution of the $\mu$-jet-jet invariant mass 
is shown in the left plot of Figure~\ref{figchi01}.
A clear peak is visible corresponding to the $\chione$
mass superimposed to a combinatorial background of events where one of the
two jets from the $\chione$ was lost and a jet from initial state radiation
was picked up. The combinatorial background can be evaluated
using three-jet events, where at least one jet is guaranteed to come
from initial state radiation. The shape of the combinatorial background estimated 
with this method is shown as the
shaded histogram superimposed to the signal peak. 
After background subtraction, an approximately gaussian  peak 
with a width of $\sim 4.5$~GeV, and a statistics of about 1050 events
is reconstructed, shown in the right of Figure~\ref{figchi01}. 
If we consider a window of $\pm12$~GeV around the peak,
corresponding  to $\sim2.5\sigma$  of the gaussian, $\sim1500$ events are observed
in the sample, and the combinatorial contamination is approximately $\sim30\%$.
A tail towards low mass values is observed, corresponding to events
where a fraction of the parton energy is lost in the jet reconstruction.
From this distribution the $\chione$ mass can be measured 
with a statistical error of $\sim 100$~MeV.
The measurement error will in this case be dominated by the systematic
error on the jet energy scale which in ATLAS is estimated to be at the level
of $1\%$ \cite{TDR}.\\ 
The $30\%$ combinatorial background is due to 
the 'soft' kinematics of the chosen example point, with a $\chione$ which is both
light and produced with a small boost. In order to show the effect of the
mass hierarchy of the involved sparticles,
the shape of the $\chione$ mass peak is shown in Figure~\ref{figchi} for a sneutrino 
mass of 500~GeV and different choices for the $\chione$ mass. In all cases 
the $\chionepm$ mass is twice the $\chione$ mass, corresponding to 
the gauge unification condition and to $|\mu|$ values of 
the same order as $M_2$. The combinatorial
background is in general smaller than for a 300~GeV sneutrino, due to the higher boost
imparted to the $\chione$, and it decreases  with increasing 
$\chione$ masses, due to the 
higher efficiency for reconstructing both jets from the $\chione$ decay. 
For this analysis no attempt has been done for the recalibration of the
jet energy. This results in the skewing of the distributions towards 
low masses, and in the peak value being slightly displaced 
with respect to the nominal mass value.\\
Once the position of the $\chione$ mass peak is known, 
the reconstructed $\chione$ statistics 
can be increased  by also considering signatures 
2) and 3-4).
For events coming from signatures 2 to 4, the $\chione$ candidate is 
defined as the muon-jet-jet combination which gives a mass 
nearest to the mass peak determined from signature 1) events. In all cases 
the reconstructed mass is required to be within $\pm12$~GeV of the peak position
to define a $\chione$ candidate. 
In $83\%$ of the events containing at least a combination satisfying 
this requirement,  only one $\chione$ candidate is found,
and this sample can be used
to improve the statistical precision on the $\chione$ mass measurement.\par
%
%--------------  FIGURE 6 : Mass chi0-W  --------------------------
\begin{figure}[htb]
\begin{center}
\dofig{8cm}{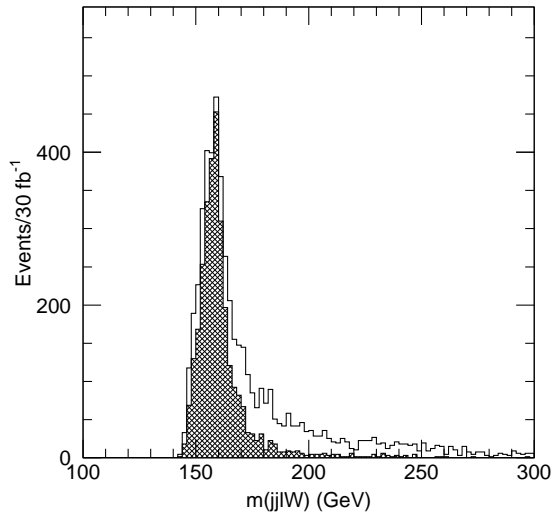}
\protect\caption{\em Invariant mass of the $\chione$ with the W candidate. The full
line histogram includes both solutions for the neutrino longitudinal
momentum, the grey one only includes the solution which gives the mass
nearest
to the measured peak.}
\label{figchip}
\end{center}
\end{figure}
%-------------------------------------------------------------------
%
%
Using the above definition of the $\chione$, we can go further in 
the mass reconstruction of the involved sparticles.
Only configurations 1) and 2) are used,
i.e. the events containing two muons and an electron 
in order to avoid ambiguities in the choice of the lepton from the $W$ decay.
The preliminary step for the reconstruction of the $\chionepm$ is 
the reconstruction of the $W$ boson from its leptonic decay.
The longitudinal momentum of the neutrino from the $W$ decay is calculated from the
missing transverse momentum of the event (considered as $p_T^{\nu}$) and the
requirement that the electron-neutrino invariant mass gives the $W$ mass.
The resulting neutrino longitudinal momentum, 
has a twofold ambiguity. We therefore build the invariant $W-\chione$ mass 
candidate using both solutions for the $W$ boson momentum.
The resulting spectrum is shown in Figure~\ref{figchip},
as the full line histogram.
A clear peak is seen, superimposed on a combinatorial background. If only
the solution yielding the  $\chionepm$ 
mass nearest to the measured mass peak is retained, 
the mass spectrum corresponding to the shaded histogram is obtained.
The peak in the unbiased histogram can be fitted with a gaussian shape, 
with a width of \mbox{$\sim6$~GeV}. \\
The combination with the mass nearest to the measured peak is
taken as $\chionepm$ candidate,  provided that the reconstructed
mass is within 15~GeV of the peak. For $80\%$ of the $e \mu \mu$ events where a 
$\chione$ candidate is found, a $W-\chione$ combination satisfying 
this requirement is reconstructed.\par
%
%--------------  FIGURE 7 : Mass chip-lepton ------------------------
\begin{figure}[t]
\begin{center}
\dofig{8cm}{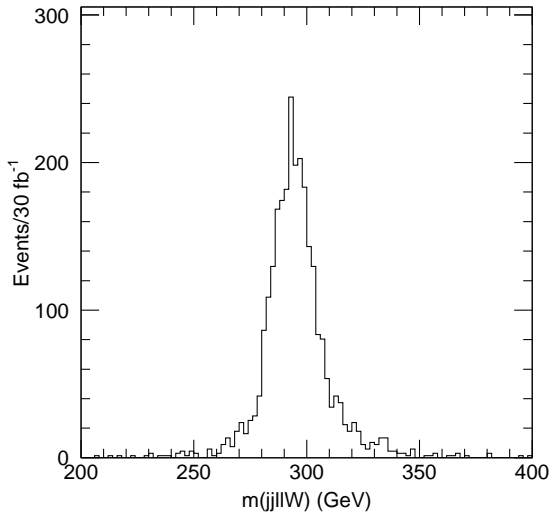}
\protect\caption{\em Invariant mass of the third lepton in the event with the
$\chionepm$ candidate.}
\label{figsl}
\end{center}
\end{figure}
%-------------------------------------------------------------------
%
Finally the $\chionepm$  candidates  are combined with the leftover muon, 
yielding the mass spectrum shown in Figure~\ref{figsl}. The $\tilde\nu$ 
mass peak 
at this point presents very limited tails, and has a width
of \mbox{$\sim10$~GeV}. 
We define fully reconstructed events as those
for which this mass lies within 25 GeV of the measured
$\tilde{\nu}$ peak.  From the estimate of the combinatorial under
the $\chione$ peak, we expect approximately 
$2 \times 1050 = 2100$ events where all the jets
and leptons are correctly assigned over a total of 2450 events observed 
in the peak. The difference between the two numbers are events for which 
one of the two jets used for the $\chione$ reconstruction comes 
from initial state radiation. These jets are typically soft, and therefore
the reconstructed $\chione$ candidate very often has a momentum which both in magnitude
and direction is close to the momentum of the original $\chione$.
Therefore for such events the  reconstructed $\chione$ behaves in the
further steps in the reconstruction as the real one, only inducing 
some widening in the $\chionepm$ and $\tilde\nu$ peaks.
\\
The statistics available at the different steps in the analysis for an integrated
luminosity of 30~fb$^{-1}$ is given in the first column of Table~\ref{tbg}. 
%
%--------------------   TABLE 1
\begin{table}
\begin{center}
\begin{tabular}{|l|r|r|r|r|r|r|}
\hline
Process & Signal & $\bar tt$ & $WZ$ & $Wbb$ & $Wt$ & $Zb$ \\
\hline
$\sigma$ (pb) & 3.3 & 590 & 26 & 300 & 60 & 7000 \\
N$_{ev}$(30 fb$^{-1}$) & $1\times 10^5$ & $1.7\times 10^7$ & $8\times 10^5$ & $9\times 10^6$ &
$1.8 \times 10^6$ & $2.1 \times 10^8$ \\
Loose cuts    & 23600 & 2900 & 53   & 2.4 & 3.5 & 56 \\
Jet veto      & 14200 & 1450 & 38   &  -  &  -  & 30  \\
$\chione$     &  6750 & 158  &  4   &  -  &  -  & -   \\
$\chionepm$   &  2700 &  8   &  0.4 &  -  &  -  & -   \\
$\snm$        &  2450 &  0   & 0.25 &  -  &  -  & -  \\
\hline
\end{tabular}
\protect\caption{\em Cross-sections and expected numbers of events after cuts
for the signal and the different Standard Model
background contributions considered in the analysis.
The ``Loose cuts" are described at the beginning of section
\ref{secana}, and the ``Jet veto" consists in adding the requirement that no third
jet with $p_T>15$~GeV is reconstructed in the event.
The line labelled ``${\tilde{\chi}}^0_1$" gives the number of events
from signatures 1 to 4 ($e \mu \mu$ and $\mu \mu \mu$)
for which a $\tilde{\chi}^0_1$ candidate is found.
The line labelled  ``${\tilde{\chi}}^{\pm}_1$" shows the
number of events from signatures 1 and 2 ($e \mu \mu$)
where a $\tilde{\chi}^{\pm}_1$ candidate is found in
addition, and the last line indicates the number of
fully reconstructed events.
In the case of the signal, we give the cross-section for the resonant
sneutrino production multiplied by the branching ratios into three leptons.
}
\label{tbg}
\end{center}
\end{table}
%--------------------------------------------------------------------------
%
For the assumed value of the coupling, $\lambda^{\prime}_{211}=0.09$, 
the uncertainty on the measurement of all the three masses involved will be 
dominated by the 1$\%$ uncertainty on the jet energy scale.\\
The efficiency for the reconstruction of the full decay chain 
with the analysis described above is \mbox{$\sim2.5\%$}. A more sophisticated analysis 
using also the three-muons events should approximately double this efficiency.\\
From the observed number of events and the 
$\tilde\nu$ mass a measurement of the quantity
$\lambda^{\prime 2}_{211}\times~BR$, where $BR$ is the product of the branching ratios
of the decays shown in equation \ref{eqq}, is possible. 
The measurement of additional SUSY processes 
will be needed to disentangle the two terms of this product. 
%
%
%========================================
\subsection{Standard Model Background}
%========================================
%
\label{secback}
The requirement of three isolated leptons in the events 
strongly reduces the possible background sources.  
The following processes were considered as a background:
\begin{itemize}
\item
$\bar tt$ production, followed by $t\rightarrow Wb$, where the two $W$ and one 
of the $b$ quarks decay leptonically. 
\item
$WZ$ production, where both bosons decay leptonically. 
\item
$Wt$ production
\item
$Wbb$ production
\item 
$Zb$ production
\end{itemize}
These backgrounds were generated with the PYTHIA MonteCarlo~\cite{PYTHIA}, except 
$Wt$ and $Wbb$
for which the ONETOP parton level generator~\cite{ONETOP} was used, 
interfaced to PYTHIA for 
hadronisation and fragmentation.
The cross-sections for the various processes, and the number of
total expected events 
for an integrated luminosity of 30~fb$^{-1}$ 
are given in Table~\ref{tbg}, according to the cross-section numbers used in 
the ATLAS physics performance TDR \cite{TDR}. 
In particular, even when the cross-section 
is known at NLO, as in the case of the top, the 
Born cross-section is taken for internal
consistency of the study.\\
For each of the background processes a sample  of events between one seventh and
a few times the expected statistics was generated and passed through the
simplified simulation of the ATLAS detector.\par       
%
%--------------- FIGURE 8 : m_chi0 : signal and SM background --------------
\begin{figure}
\begin{center}
\dofig{8cm}{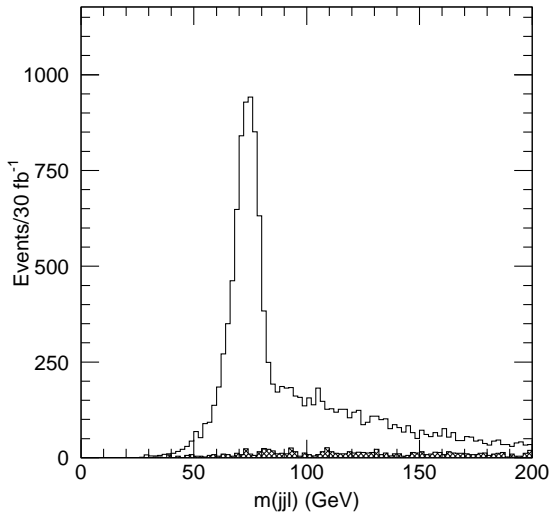}
\protect\caption{\em Invariant  mass of the $\chione$ candidates entering the 
kinematic analysis superimposed to the 
Standard Model background (hatched).}
\label{figbg}
\end{center}
\end{figure}
%---------------------------------------------------------------------------
%
After the loose selection cuts described in Section~\ref{secana}, 
the background is dominated
by top production, as can be seen from the numbers shown in Table~\ref{tbg}. 
The distribution of the $\mu$-jet-jet invariant mass for 
background events, obtained as in Section \ref{secana} 
and corresponding to the $\chione$ candidates selection, 
is shown as the hatched histogram in Figure~\ref{figbg}.
In this figure we have superimposed the same distribution for the signal. 
Already at this level, the signal stands out very clearly from the background, and in the
following steps of the reconstruction the background becomes almost
negligible. The numbers of 
background and signal events expected at the various steps of the reconstruction
can be compared in Table~\ref{tbg}. 
The full analysis was performed only for the $\bar tt$ and
$WZ$ background because for the other channels the background is essentially 
negligible compared to top production, and  in most cases the MonteCarlo statistics
after the initial selection was too low to allow a detailed study. 
For the SUSY model considered and the chosen value of the $\lambda^{\prime}$
coupling constant, even the loose selection applied allows to efficiently
separate the signal from the background. 
%
%============================================
\subsection{Sensitivity on $\lambda^\prime$}
\label{seclam}
%============================================
%
From these results, it is possible to evaluate the minimum value of the 
$\lambda_{211}^{\prime}$ coupling for which it will be possible to discover the signal.
The starting point in the analysis is the observation of a peak in the 
muon-jet-jet invariant mass
over an essentially flat background. All of the further analysis steps
of the cascade reconstruction rely on the possibility of selecting the
events with a mass around the  $\chione$ peak.\\
For the observation of the peak, the best signal/background ratio is obtained using the 
three-muons sample (configurations 3 and 4 above). 
In the Standard Model, which incorporates lepton universality, 
about one eight of the
three-leptons  events present a three-muons configuration, 
whereas about half of the signal 
events come in this configuration, thereby granting an improvement of a factor 4 
in signal over background, with respect to the full sample.
The three muons come either in the  '$-++$' or in the '$-+-$' sign configuration,
because the two muons from the decay chain
$\tilde\nu (\overline{\tilde\nu} )\to\chionepm\mu^{\mp}\to\chione\mu^{\pm}\mu^{\mp}$ must
have opposite sign, whereas the $\chione$ can decay to muons of either sign.
Therefore the muon for the $\chione$ reconstruction must be chosen 
between the two same-sign ones.
%
%-----------  FIGURE 9 : M_chi0 for 3 muon events -----------------------
\begin{figure}[htb]
\begin{center}
\dofig{8cm}{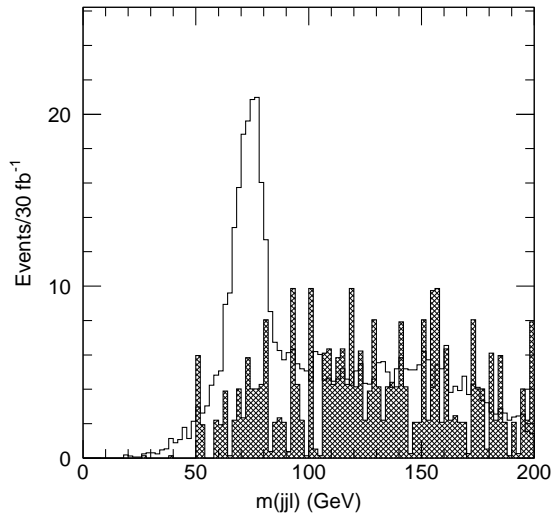}
\protect\caption{\em Invariant  mass of the $\chione$ candidates from three-muon
events scaled down by a factor 25,
corresponding to a $\lambda^{\prime}$ value of 0.018,
superimposed to the
Standard Model background (hatched).}
\label{figbg1}
\end{center}
\end{figure}
%------------------------------------------------------------------------
%
The distribution for the $\mu$-jet-jet invariant mass, for events containing two
jets and three muons is shown in Figure~\ref{figbg1}, scaled down 
by a factor 25,  corresponding to a $\lambda^{\prime}$ value of 0.018, 
superimposed to the expected  top background. In the distribution
each event enters twice, for each of the two same-sign muons which can be 
used  to reconstruct the $\chione$.
We expect, however,  that the
combination with the ``wrong" muon gives in most cases a reconstructed
mass outside of the $\chione$ peak.\\
A statistical prescription is needed to 
define the fact that a peak structure is seen in the signal+background distribution.
Given the exploratory nature of the work, we adopt the naive approach
of calculating the $\lambda^\prime$ value for which $S/\sqrt{B}=5$,
where $S$ and $B$ are respectively the number of signal and background candidates 
counted in an interval of $\pm 15$~GeV around the measured $\chione$ peak. 
The window for the definition of a $\chione$ candidate is enlarged with
respect to the analysis described in Section~\ref{secana}, 
in order to recover the non-gaussian tail of the signal peak, thus increasing
the analysis efficiency.
In this interval, for the chosen point, for an integrated luminosity of 30~fb$^{-1}$, 
$S=580000\times (\lambda^\prime)^2$ and $B=46$ events. In the hypothesis that
the $\bar tt$ background can be precisely measured from data, the 
lower limit on $\lambda^\prime_{211}$ is:
$$\lambda^\prime_{211}>0.0075$$
\\
The pair production of SUSY particles through standard $R_p$-conserving
processes is another possible source of background, due to the possibility 
to obtain final states with high lepton multiplicity, and the high production cross-sections.
This background 
can only be evaluated inside  models  providing predictions 
for the whole SUSY spectrum. As a preliminary study, a sample of 
events were generated with the HERWIG 6.0 MonteCarlo~\cite{herwig} by setting the 
slepton masses at 300~GeV, the masses of squarks and gluinos at 1000~GeV
and the chargino-neutralino spectrum as for the example model.
The total $R_p$-conserving cross-section is in this case \mbox{$\sim6$~pb}.
A total of 60 SUSY background events which satisfy the requirements 
used above to define $S$ and $B$ are observed.
All the events surviving the 
cuts are from direct chargino and neutralino
production, with a small contribution
from Drell-Yan slepton production. Since the contributions 
from squark and gluino decays are strongly suppressed by the jet veto requirements,
this result can be considered as a correct order of 
magnitude estimate, independently 
from the assumed values for the squark and gluino masses.
Moreover, the reconstructions of the chargino and sneutrino masses
can also be used in order to reduce the SUSY background.  
A more thorough discussion of the SUSY  
background 
will be given below in the framework of the mSUGRA model.
%
%
%
%%%%%%%%%%%%%%%%%%%%%%%%%%%%%%%%%%%%%%%%%%%%
\section{Analysis reach in various models}
\label{reach}
%%%%%%%%%%%%%%%%%%%%%%%%%%%%%%%%%%%%%%%%%%%%%
%
For the example case studied in Section~\ref{secana} it was shown  
that the sneutrino production 
signal can be easily separated from the background, and allows to 
perform precision measurements of the masses of the sparticles 
involved in the decay chain. \\
The analysis can be generalised to investigate the range of SUSY 
parameters in which this kind of analysis is possible, and 
to define the minimum value of the $\lambda^\prime$ constant
which gives a detectable signal in a given SUSY scenario.
The different model parameters enter the definition of the detectability 
at different levels:
\begin{itemize}
\item
The sneutrino production cross-section is a function only of the
sneutrino mass and of the square of the R-parity violating coupling constant.
\item
The branching fraction of the sneutrino decay into three 
leptons is a function of all the SUSY parameters. 
\item
The analysis efficiency is a function of the 
masses of the three supersymmetric particles involved in the decay.
\end{itemize} 
%
%----------  FIGURE 10 : Efficiency vs m_chip -----------------------------
\begin{figure}[htb]
\begin{center}
\dofig{8cm}{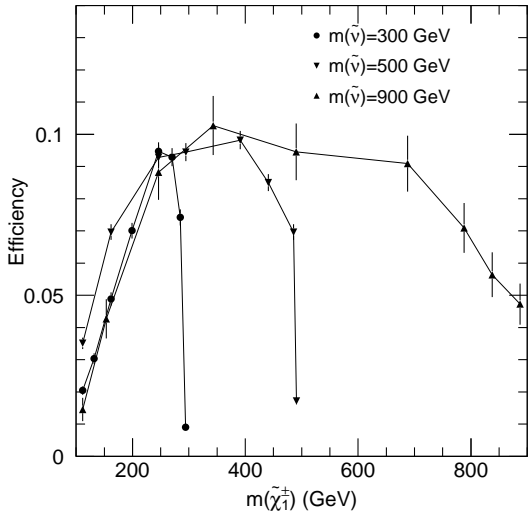}
\protect\caption{\em Efficiency for reconstructing  a $\mu$-jet-jet
invariant mass within 15 GeV of the $\chione$ mass in three-muons events,
as a function of the $\chionepm$ mass. The shown points were generated 
with the following parameters: \mbox{$-\mu=M_2=2M_1$}, \mbox{$\tan\beta=1.5$}. 
All the sfermion masses are set equal to the sneutrino mass.
Points  for $m_{\tilde\nu}= 300, 500$ and $900$~GeV are shown.}
\label{pleff}
\end{center}
\end{figure}
%----------------------------------------------------------------------------
%
The dependences of the cross-section and branching ratios 
on the SUSY parameters were discussed in Section \ref{secmssm} for the MSSM, 
and are summarised in Figures~\ref{mssmu} and \ref{mssmxse}.
We only need at this point to parameterize the analysis efficiency
as a function of the sparticle masses. 
The number of signal events for each considered model will then be obtained 
by multiplying the expected number of three-lepton events by the
parameterized efficiency.
\subsection{Efficiency of the three-muon analysis}
According to the discussion presented 
in Section~\ref{seclam}, we need to calculate 
the efficiency for the signal process to satisfy the following requirements:
\begin{itemize}
\item
to pass the initial selection cuts described in Section~\ref{secana}
(Loose cuts), including
the veto on the third jet (Jet veto);
\item
to contain three reconstructed muons, with one of the $\mu$-jet-jet invariant masses 
within 15 GeV of the $\chione$ mass.
\end{itemize}
Three sneutrino  masses, $m_{\tilde\nu}=~300, 500$ and $900$~GeV
were considered, and for each of these the evolution of the efficiency 
with the $\chionepm$ mass was studied.  The mass of the $\chione$ was 
assumed to be half of the mass of the $\chionepm$, relation 
which is in general valid in SUGRA inspired models and correspond
to a choice of values for $|\mu|$ of the same order as $M_2$. \\
The analysis efficiency 
is shown in Figure~\ref{pleff} as a function of the $\tilde\nu_{\mu}$ mass and
of the $\chionepm$ mass. The efficiency values are 
calculated with respect to the number of events 
which at generation level did contain the three leptons, therefore 
they only depend on the event kinematics and not on the
branching ratios.
The loss of efficiency at the lower end of the  $\chionepm$
mass spectrum is due to the inefficiency for detecting two jets
from the $\chione$ decay, either because the two jets are 
reconstructed as a single jet, or because one of the two jets 
is below the detection threshold of 15 GeV. 
The efficiency then becomes approximately independent of the masses of
the sneutrino and of the $\chionepm$, up to the point
where the $\tilde\nu$ and $\chionepm$ masses become close enough to affect 
the efficiency for the detection of the muon from the 
\mbox{$\tilde\nu\to\chionepm\mu$} decay; for $m_{\tilde\nu}-m_{\chionepm}<10$~GeV
the analysis efficiency rapidly drops to zero. The moderate decrease 
in efficiency at high $\chionepm$
masses for $m_{\tilde\nu}=900$~GeV can be ascribed to the fact that one of two
energetic jets from the $\chione$ decay radiates a hard gluon, three jets 
are reconstructed, and the event is rejected by the jet veto.
\\
At this point all the ingredients are available  to study 
the reach in the parameter space for the analysis presented in Section~\ref{analysis}
within different  SUSY models. \\
%
%=================
\subsection{Analysis reach in the MSSM}
\label{mssm}
\begin{figure}[h]
\begin{center}
\dofig{9cm}{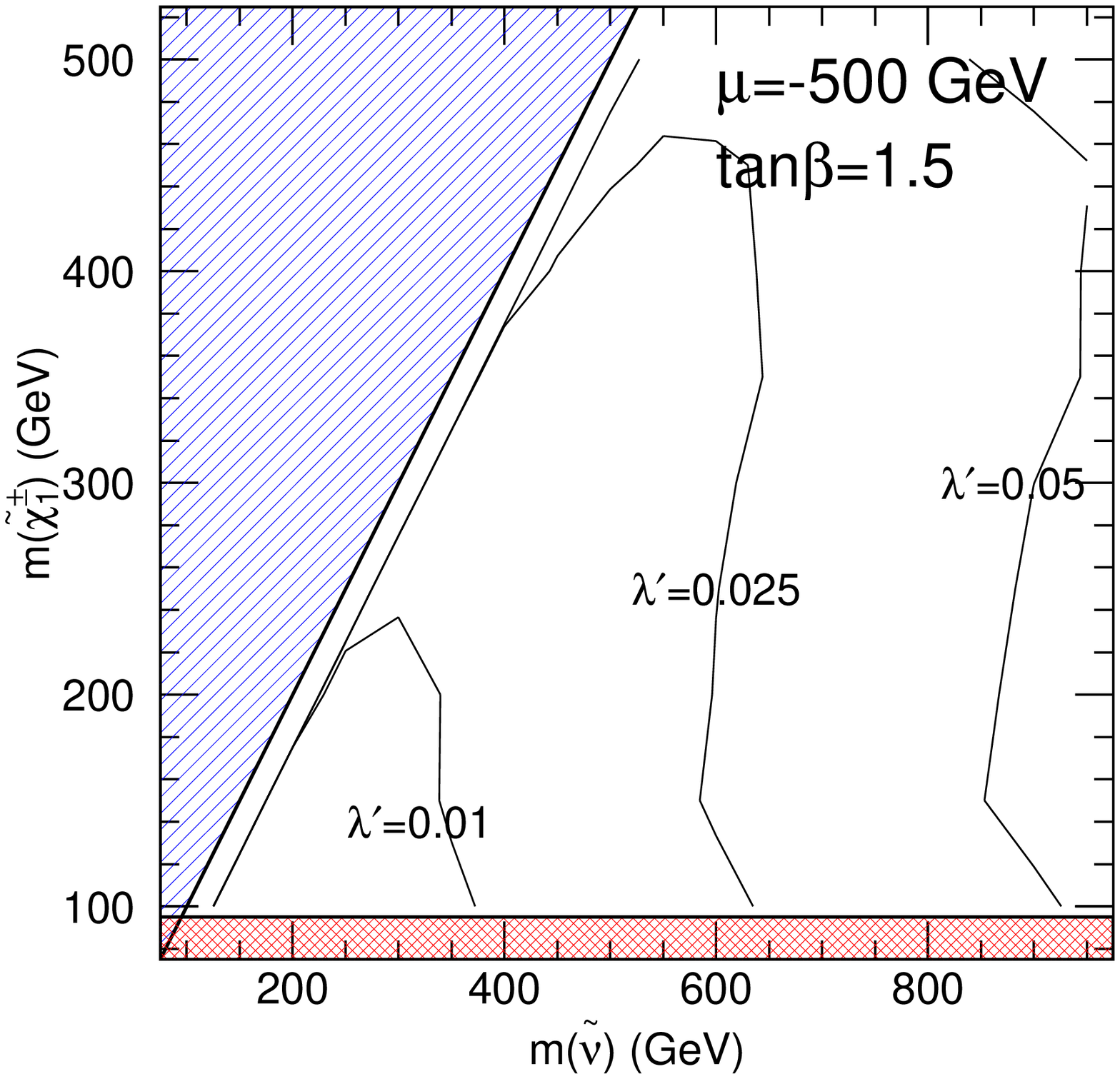}
\protect\caption{\em
$5\sigma$ reach in the $m_{\tilde\nu}$-$m_{\chionepm}$
plane for three
different choices of the $\lambda^\prime_{211}$ coupling for 
an integrated luminosity of 30~fb$^{-1}$ at the LHC.
The chosen model parameters were:
\mbox{$\mu=-500$~GeV}, \mbox{$\tan\beta$=1.5}, 
\mbox{$m_{\tilde q}=m_{\tilde l}=300$~GeV},
\mbox{$A_t=A_b=A_{\tau}=0$}, \mbox{$M_2=2M_1$}.
The significance is defined
only considering the Standard Model background, and a signal
of at least ten events is required. 
The hatched region at the upper
left corresponds to \mbox{$m_{\tilde \nu}<m_{\tilde\chi^{\pm}_1}$}.
The cross-hatched region at low $m_{\chionepm}$ is excluded by the
preliminary LEP results at $\sqrt{s}=196$~GeV \cite{aleph2}.}
\label{mssmreach}
\end{center}
\end{figure}

%=================
%
The region in the $m_{\tilde\nu}$-$m_{\chionepm}$ plane for which the
signal significance is  greater than 5$\sigma$, as defined  in Section~\ref{seclam},
and at least 10 signal events
are observed for an integrated luminosity of
30~fb$^{-1}$ is shown in Figure~\ref{mssmreach}
for different choices of the $\lambda^{\prime}_{211}$ constant.
%
%---------   FIGURE 11 : 5 sigma reach in MSSM, (m_chip, m_snu) plane
%-------------------------------------------------------------------------
%
The behaviours of the sensitivity curves in the $m_{\tilde\nu}$-$m_{\chionepm}$ plane
are well explained by the variations of the single chargino production 
cross-section shown in Figure \ref{mssmxse} in the same plane of parameters.   
The SUSY background is not considered in the plot, as it depends on 
all the model parameters. It was however verified in a few example cases that 
for our analysis cuts this background is dominated 
by direct chargino and neutralino production, and  it
becomes negligible in the limit of high 
$\chione$ and $\tilde \chi^{\pm}$ masses. The main effect
of taking into account this background will be to reduce the significance 
of the signal for $\chionepm$ masses lower  than 200~GeV.\par
From the curves in Figure~\ref{mssmreach} we can conclude that  
within the MSSM, the production of
a 900~GeV sneutrino for $\lambda^{\prime}_{211}>0.05$, and of a 350~GeV sneutrino for
$\lambda^{\prime}_{211}>0.01$ can be observed 
within the first three years of LHC running,
provided that the sneutrino is heavier than 
the lightest chargino.\par
The sensitivity on an \rpv coupling of type $\l'_{2jk}$ can be derived from 
the sensitivity obtained for $\l'_{211}$, as explained in Section~\ref{xsec}.
For example, we have seen that the sensitivity on $\l'_{221}$ was
$\sim 1.5$ times weaker than the sensitivity on $\l'_{211}$,
for $\tan\beta$=1.5, $M_2=200$~GeV, $\mu=-500$~GeV and $m_{\tilde \nu}=400$~GeV.
This set of parameters leads to a sensitivity on 
$\l'_{211}$ of about $0.015$ as can be seen in Figure~\ref{mssmreach},
and hence to a sensitivity on $\l'_{221}$ of $\sim 0.022$.  
In Table \ref{couplg}, we present the sensitivity on any $\l'_{2jk}$ coupling 
estimated using the same method and for the same MSSM parameters. 
Those sensitivities represent an important
improvement with respect to the low-energy limits of \cite{Drein}. \\
In the case of a single dominant $\l'_{2j3}$ coupling 
the neutralino decays as $\tilde \chi^0_1 \to \mu u_j b$
and the semileptonic decay of the b-quark could affect
the analysis efficiency. Hence in this case, the precise sensitivity
cannot be simply calculated by scaling the value obtained for
$\lambda^\prime_{211}$. The order of magnitude of the sensitivity
which can be inferred from our analysis should however be correct.  
%
%--------------------   TABLE 2
\begin{table}
\begin{center}
\begin{tabular}{|c|c|c|c|c|c|c|c|c|}
\hline
$\l'_{211}$ & $\l'_{212}$ & $\l'_{213}$ & $\l'_{221}$ & $\l'_{222}$ 
& $\l'_{223}$ & $\l'_{231}$ & $\l'_{232}$ & $\l'_{233}$  \\
\hline
0.01 & 0.02 & 0.02 & 0.02 & 0.03 & 0.05 & 0.03 & 0.06 & 0.09 \\
\hline
\end{tabular}
\protect\caption{\em Sensitivities on the $\l'_{2jk}$ coupling constants 
deduced from the sensitivity on $\l'_{211}$ for $\tan\beta$=1.5, 
$M_1=100$~GeV, $M_2=200$~GeV, $\mu=-500$~GeV,
$m_{\tilde q}=m_{\tilde l}=300$~GeV and $m_{\tilde \nu}=400$~GeV.}
\label{couplg}
\end{center}
\end{table}
%--------------------------------------------------------------------------
%
%=======================
\subsection{Analysis reach in mSUGRA}
\label{msugra}
%=======================
%
%-----------  FIGURE 12 : Isomasses for snu and chip -------------------
\begin{figure}
\begin{center}
\dofig{9cm}{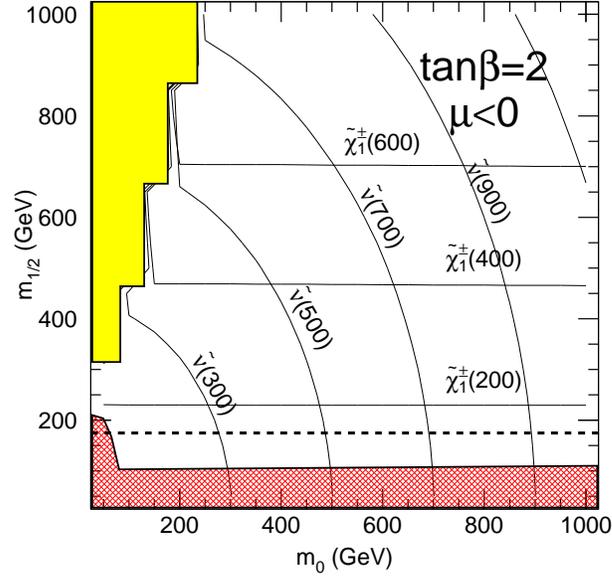}
\dofig{9cm}{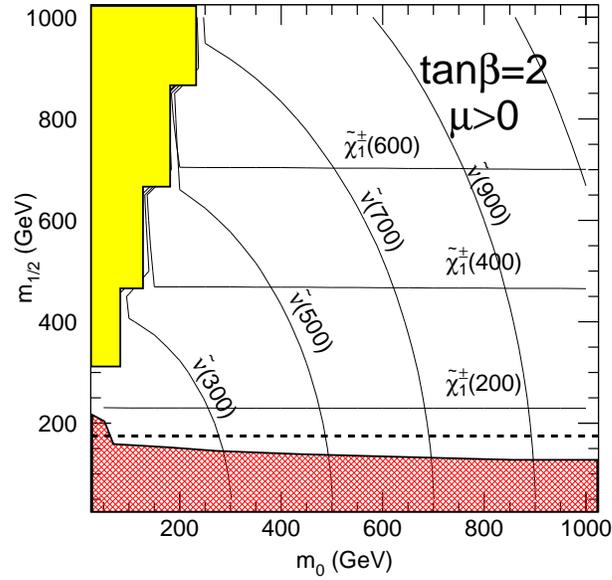}
\protect\caption{\em Curves of equal mass for $\tilde\nu$ and $\chionepm$
in the $m_0-m_{1/2}$ plane for $\tan\beta=2$.
The grey region at the upper left indicates the domain where the $\chione$ is
not the LSP.
The cross-hatched region for low $m_{1/2}$ gives the kinematic
limit for the discovery of $\chionepm$  or $\tilde l$ by LEP
running at  $\sqrt{s}=200$~GeV.
The dotted line shows the
region below which the $\chionepm$ decays to a virtual W.
}
\label{plmass}
\end{center}
\end{figure}
%------------------------------------------------------------------------
%
Our framework throughout this Section will be 
the so-called minimal supergravity model.
In this model the parameters obey a set of boundary conditions at the
Grand Unification Theory (GUT) scale $M_x$. These conditions appear
to be natural in supergravity scenario since the supersymmetry breaking 
occurs in an hidden sector which communicates with the visible sector
only through gravitational interactions. 
First, mSUGRA contains the gauge coupling unification at $M_x$, 
such an unification being suggested by the
experimental results obtained at LEP I.
One can view the gauge coupling unification assumption 
as a fixing of the GUT scale $M_x$. 
Second, the gaugino (bino, wino and gluino) masses at $M_x$ 
are given by the universal mass $m_{1/2}$.
the parameters $m_{1/2}$ and $M_i \ [i=1,2,3]$ 
are thus related by the solutions of the renormalization group 
equations (RGE). Besides, since the gaugino masses and 
the gauge couplings are governed by the
same RGE, one has the well-known relation: 
$M_1={5 \over 3} \tan^2 \theta_W M_2$.
Similarly, at $M_x$, the universal scalars mass is $m_0$ and the 
trilinear couplings are all equal to $A_0$. Finally, in mSUGRA the
absolute value of the higgsino mixing parameter 
$\vert \mu \vert$ as well as the bilinear coupling $B$ are determined by the 
radiative electroweak symmetry breaking conditions.
Therefore, mSUGRA contains only the five following parameters:
$sign(\mu)$, $\tan\beta$, $A_0$, $m_0$ and $m_{1/2}$.\par 
Due to the small  dependence of the single chargino production rate
on the $\mu$ parameter for $M_2\le|\mu|$ (see Section \ref{secmssm}), 
the study of the mSUGRA model in which $\vert \mu \vert$ is fixed 
by the electroweak symmetry breaking condition provides information on 
a broader class of models. 
The single chargino production rate depends mainly on the values 
of $m_0$ and $m_{1/2}$ (see Section \ref{secmssm}).
We will set $A_0=0$,  and study the detectability
of the signal in the $m_0-m_{1/2}$  plane. 
We show in Figure~\ref{plmass} the curves of equal mass for
$\tilde\nu$ and $\chionepm$ for $\tan\beta=2$ calculated with
the ISASUSY \cite{isajet} package which uses one-loop  RGE
to get the SUSY spectrum from the mSUGRA parameters.
\par
%------------------------------------------------------------------------
%
The signal reach can be easily evaluated from the sparticle mass
spectrum and branching fractions by using the parameterization of the
analysis efficiency shown in Figure~\ref{pleff}.
%
%----------------------------------------
\subsubsection{Supersymmetric background}
%----------------------------------------
%
In the case of a well constrained model as mSUGRA, the SUSY background
can in principle be evaluated in each considered point in the parameter space.
For this evaluation the full SUSY sample must be 
generated for each point, requiring 
the generation of a large number of events. \par
The sparticle masses for the model studied in detail in Section~\ref{analysis}
uniquely define a model in the mSUGRA  framework. Therefore, as a first approach
to the problem, a full analysis was performed for this model,  
corresponding to the parameter values: 
$$
m_{0}=275~{\mathrm{GeV}},~~~~m_{1/2}=185~{\mathrm{GeV}},~~~~\tan\beta=1.5,~~~~ \mu<0,~~~ A_0=0.  
$$
%
%----------   FIGURE 13 : Nb of susy bckgd evts in m_1/2, m_0 plane ------
\begin{figure}[htb]
\begin{center}
\dofig{10cm}{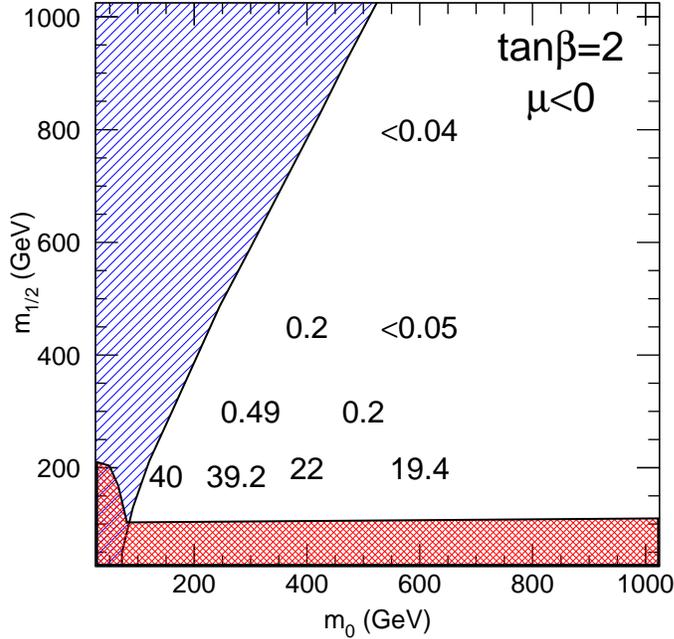}
\protect\caption{\em Number of SUSY background events for an integrated
luminosity of $30$~fb$^{-1}$ in the $m_0-m_{1/2}$ plane with $\tan\beta$=2
for a few selected test models. The hatched region at the upper
left corresponds to $m_{\tilde\nu}<m_{\tilde\chi^{\pm}_1}$.
The cross-hatched region for low $m_{1/2}$ gives the kinematic 
limit for the discovery of $\chionepm$  or $\tilde l$ by LEP 
running at  $\sqrt{s}=200$~GeV.}
\label{plback}
\end{center}
\end{figure}
%--------------------------------------------------------------------------
%
For this mSUGRA point, the mass scale for squarks/gluinos is
in the proximity of 500~GeV, and the total cross-section 
for all the SUSY particles pair productions is
approximately 130~pb, yielding a signal of \mbox{$\sim4 \ 10^6$} events for
the first three years of data-taking at the LHC.
A total of 400k events were generated and analysed.
The number of surviving events after cuts in the three-muons sample was $47\pm21$
for an integrated luminosity of 30~fb$^{-1}$. All the background events 
come from direct chargino and neutralino production, as it was the case for the 
MSSM point studied in Section~\ref{seclam}. As a cross-check, we generated
for the same mSUGRA point only the processes of the type $pp\to\tilde\chi+X$,
where $\tilde\chi$ denotes either $\tilde\chi^0$ or $\tilde\chi^{\pm}$, 
and $X$ any other SUSY particle.
The cross-section is in this case 
\mbox{$\sim6$~pb}, and the number of background events is $39\pm7$ events,
in good agreement with the number evaluated generating all the SUSY processes.\par
Based on this result, we have performed
a scan in the $m_0-m_{1/2}$ plane the fixed values $\tan\beta=2$, $\mu<0$.
On a grid of points  we generated  event samples 
for the $pp\to\tilde\chi+X$ processes with the 
HERWIG 6.0 MonteCarlo~\cite{herwig}. 
The number of SUSY events with a $\mu$-jet-jet combination with an invariant
mass within 15~GeV of the $\chione$ mass is shown in 
Figure~\ref{plback} in the $m_0-m_{1/2}$ plane for an integrated
luminosity of 30~fb$^{-1}$. The background is 
significant for a $\chionepm$ mass of 175~GeV ($m_{1/2}=$200~GeV), and becomes 
essentially negligible for $\chionepm$ mass of 260~GeV ($m_{1/2}=$300~GeV).
This behaviour is due to the combination of two effects: the $\tilde \chi^{\pm} \tilde \chi^{\pm}$
production cross-section decreases with increasing $\tilde \chi^{\pm}$ mass, and the probability
of losing two of the four jets from the decay of the two $\chione$ in the event
becomes very small for a $\chionepm$ mass of $\sim220$~GeV. Indeed,
the suppression of the SUSY background is mainly due to the Jet veto cut. \par
Given the high SUSY cross-section, and the high lepton multiplicity
from $\chione$ decays, a prominent signal should manifest itself 
through R-conserving sparticle pair production 
in this scenario. Single resonant sneutrino production will then be used as 
a way of extracting information on the value of the $R_p$-violating
coupling constant, and of precisely measuring the masses of $\tilde\nu_\mu$,
$\chionepm$, $\chione$. Moreover,
thanks to the very high number of produced $\chione$
expected from $\tilde q / \tilde g$ pair production, the
$\chione$ mass will be directly reconstructed from $\tilde q$ and
$\tilde g$ decays, as shown in \cite{TDR}. 
So, for the present analysis it can be assumed 
that the $\chione$ mass is approximately known, and an attempt to reconstruct
the $\chionepm$ peak can be performed even if the $\chione$ reconstruction
does not yield a significant peak over the SUSY+SM background.  
In order to perform the full reconstruction, one just needs 
to observe a statistically significant 
excess of events over what is expected from the Standard Model 
background in the mass region corresponding to the known $\chione$ mass.
The full kinematic reconstruction described in Section~\ref{secana} above
will then easily allow to separate the process of interest from the SUSY background.
%
%---------------------
\subsubsection{Results}
%---------------------

%---------   FIGURE 14 : 5 sigma reach in mSUGRA, low tan(beta) -------------
\begin{figure}
\begin{center}
\dofig{9cm}{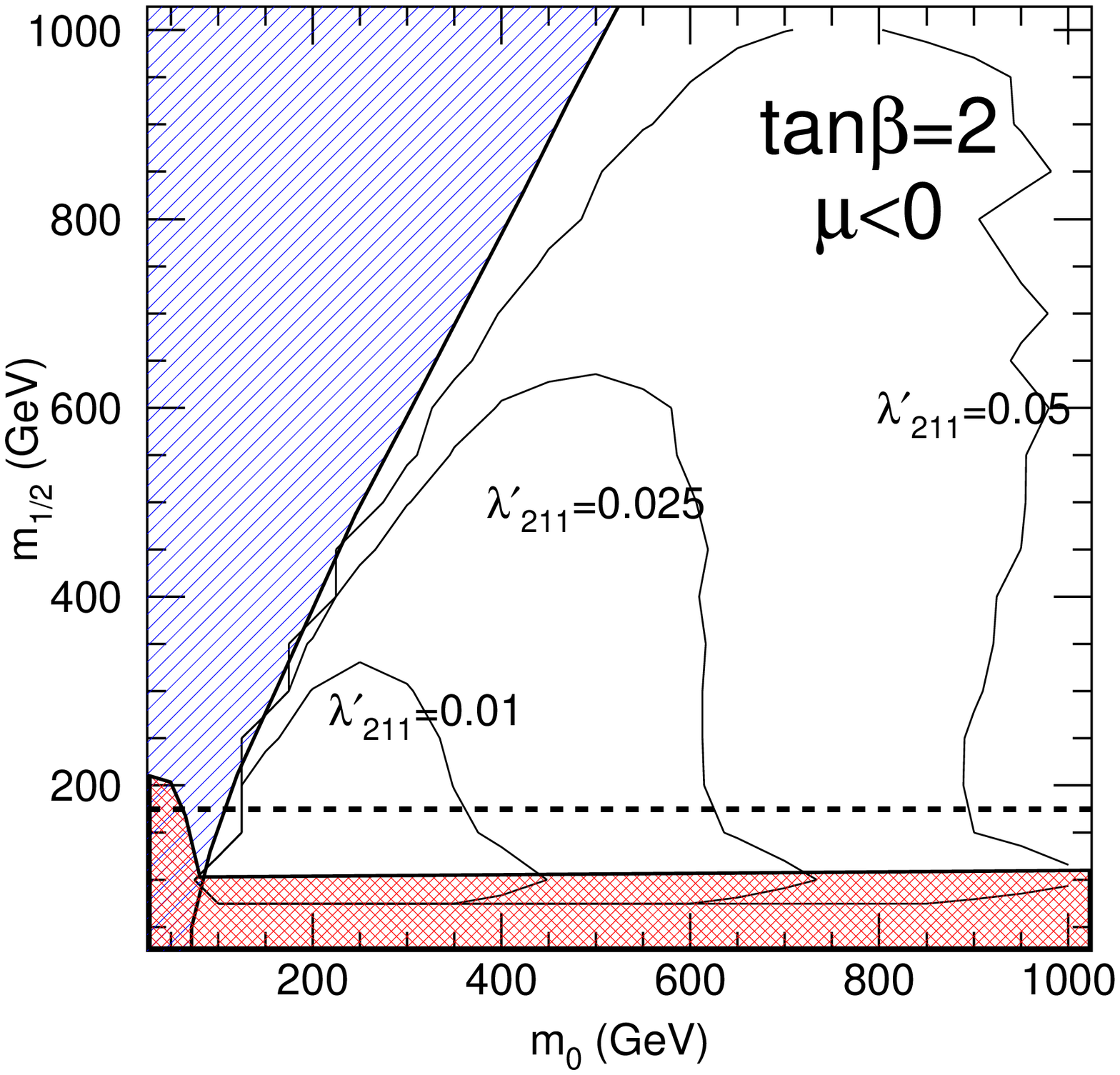}
\dofig{9cm}{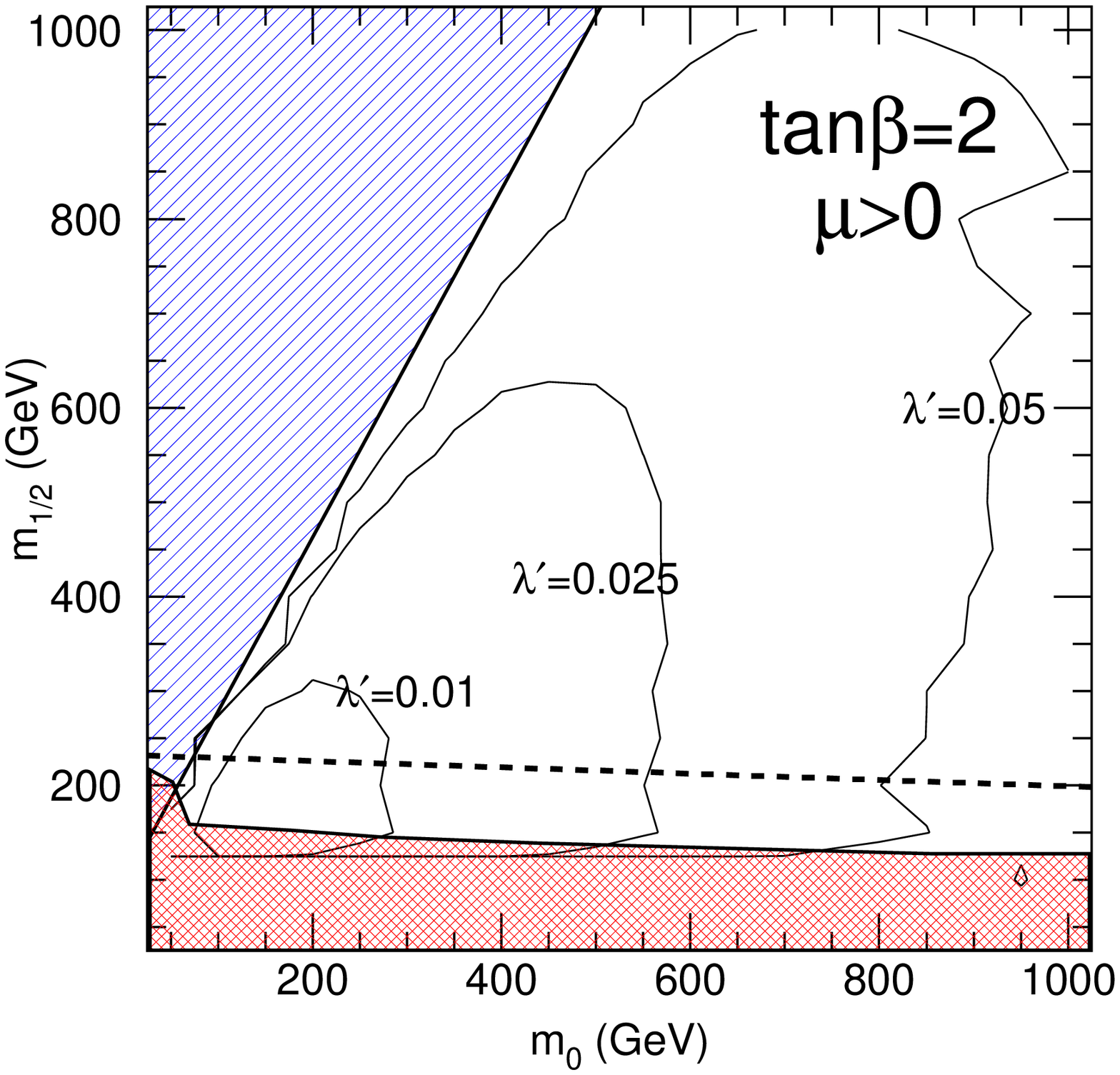}
\protect\caption{\em $5\sigma$ reach in the $m_0-m_{1/2}$ plane for $\tan\beta$=2 and three 
different choices of the $\lambda^\prime_{211}$ coupling for an integrated
luminosity of 30~fb$^{-1}$  at the LHC. The significance is defined 
only considering the Standard Model background, and a signal
of at least ten events is required. The dotted line shows the
region below which the $\chionepm$ decays to a virtual W.
}
\label{plreach}
\end{center}
\end{figure}
%------------------------------------------------------------------------

Based on the discussion in the previous section we calculate the signal significance as
$S/\sqrt{B}$, where for the
signal $S$ we only consider the resonant sneutrino production, 
and for the background $B$ we only consider the SM background.
We show in Figure~\ref{plreach} for $\tan\beta=2$ and for the two 
signs of $\mu$ the regions 
in the $m_0-m_{1/2}$ plane for which the signal significance exceeds 5$\sigma$
and the number of signal events is larger than 10, for an integrated 
luminosity of 30~fb$^{-1}$. The reach is shown for three different
choices of the $\lambda^{\prime}_{211}$ parameter: 
$\lambda^{\prime}_{211}=0.01,0.025,0.05$. Even for the lowest 
considered coupling the signal can be detected in a significant 
fraction of the parameter space. The dotted line shows the
region below which the $\chionepm$ decays to the $\chione$ and a virtual $W$, 
thus making the kinematic reconstruction of the decay chain described
in Section~\ref{secana} impossible.
The reconstruction of the $\chione$ is however still possible,
but the reconstruction efficiency  drops rapidly due to the difficulty 
to separate the two soft jets from the $\chione$ decay. A detailed study 
involving careful consideration of jet identification algorithms 
is needed to assert the LHC reach in that region. \par

%---------  FIGURE 15 : 5 sigma reach in mSUGRA, high tan(beta) ------
\begin{figure}
\begin{center}
\dofig{9cm}{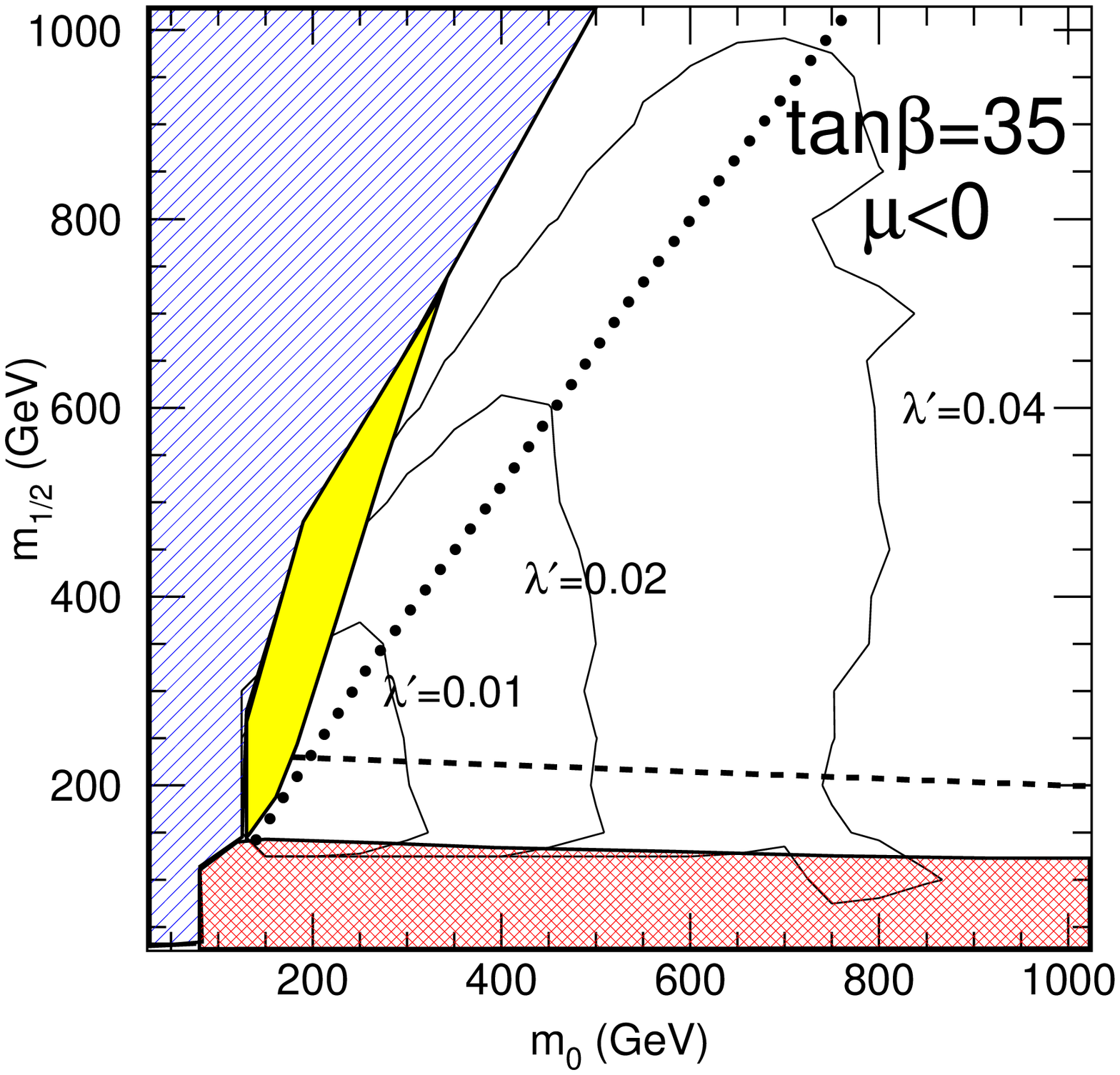}
\dofig{9cm}{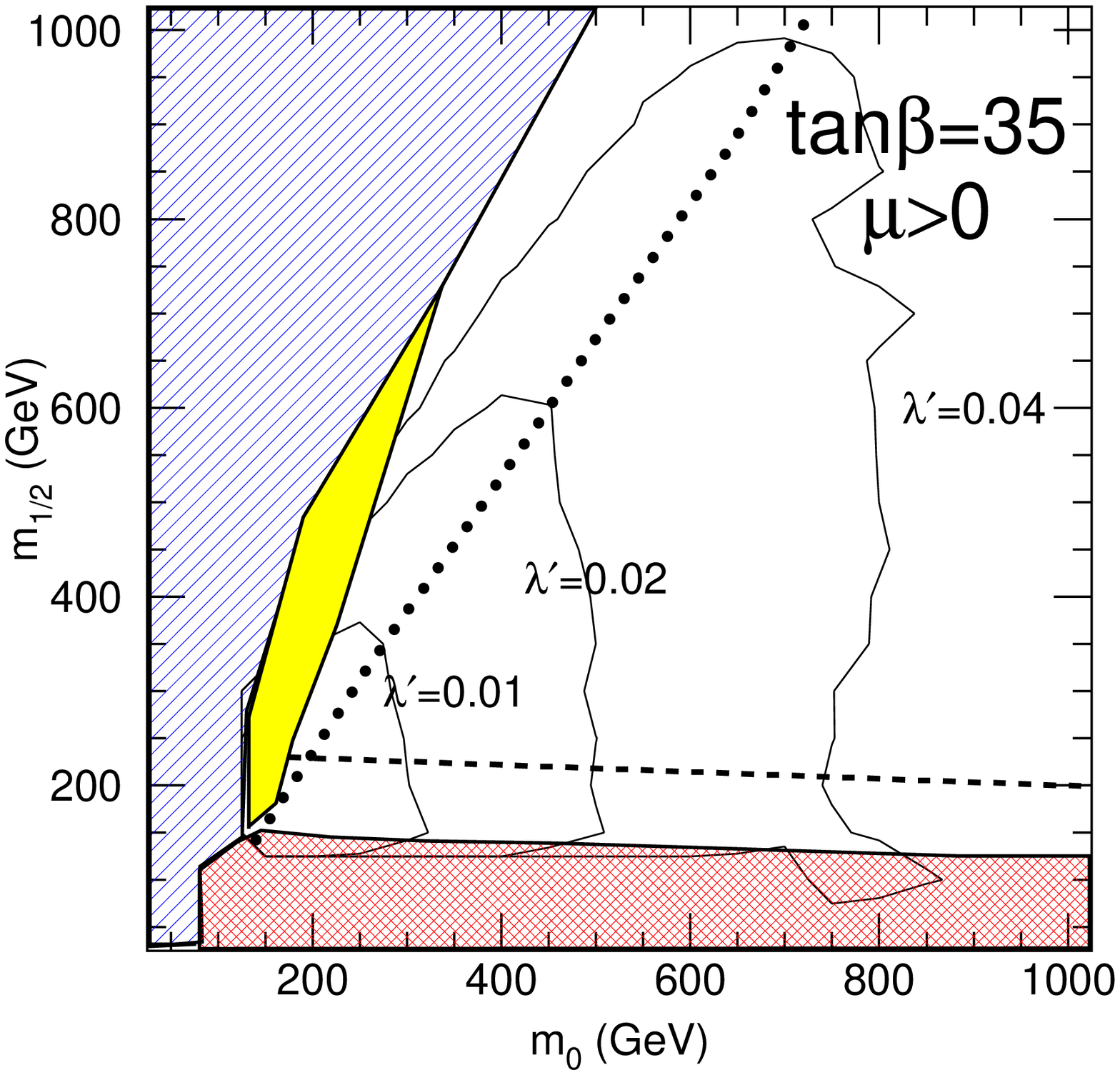}
\protect\caption{\em $5\sigma$ reach in the $m_0-m_{1/2}$ plane for $\tan\beta$=35 and three
different choices of the $\lambda^\prime_{211}$ coupling for an integrated
luminosity of 30~fb$^{-1}$  at the LHC. The significance is defined
only considering the Standard Model background, and a signal
of at least ten events is required. The dashed line shows the
region below which the $\chionepm$ decays to a virtual W.
The region to the left of the dotted lines has a branching ratio
for $\chionepm\to\tilde\tau_1\nu_{\tau}$ larger than 50$\%$,
and the grey area indicates the region for which a low signal efficiency is expected.
}
\label{plreach35}
\end{center}
\end{figure}
%------------------------------------------------------------------------

As observed in \cite{greg}, the efficiencies  quoted for this analysis 
rely  on a branching ratio of $\sim100\%$ for the decay
$\chionepm\to W\chione$. This is in general true in SUGRA models
as long as the $\tilde\tau_1$ is heavier  than the $\chionepm$, corresponding
to moderate values for $\tan\beta$.  
For high $\tan\beta$, the decay  $\chionepm\to\tilde\tau_1\nu_{\tau}$
can become kinematically possible,
and its branching ratio can dominate the standard
$\chionepm\to W\chione$. 
The stau in turns typically decays as $\tilde \tau_1 \to \tau \tilde \chi^0_1$.
The three-leptons  signature is in this case
even enhanced, due to the higher branching fraction 
into electrons and muons for the $\tau$ compared to the $W$, 
at the price of a softer lepton spectrum.
The $\chione$ reconstruction is still possible
but the presence of three neutrinos 
(two additional neutrinos come from the leptonic $\tau$ decay) 
renders the reconstruction of the particles earlier in the decay 
chain difficult. The analysis efficiency is essentially unaffected
with respect to the low $\tan\beta$ case as long as the mass difference between
the $\tilde\tau_1$ and the $\chione$ is larger than $\sim 50$~GeV.
For $\tilde\tau_1$ and $\chione$ masses too much degenerate,
the transverse momentum of the charged lepton coming from the $\tau$ 
decay would often fall below the analysis requirements,
leading thus to a reduction of the signal efficiency. 
The reach in the $m_0-m_{1/2}$ plane is shown in Figure~\ref{plreach35}
for $\tan\beta=35$ and three 
different choices of the $\lambda^\prime_{211}$ coupling.
The branching fraction for the decay $\chionepm\to\tilde\tau_1\nu_{\tau}$ is
higher than 50$\%$ to the left of the dotted line, and the region 
for which a reduced signal efficiency is expected is displayed as a grey area.
The reach for $\chione$ detection is similar 
to the low $\tan\beta$ case, 
but the region in which the full reconstruction of the 
sneutrino decay chain is possible is severely restricted.

%%%%%%%%%%%%%%%%%%%%%%%%%%%%%%%%
\section{Conclusions}
%%%%%%%%%%%%%%%%%%%%%%%%%%%%%%%%
%
We have analysed the resonant sneutrino production at LHC in supersymmetric
models with R-parity violation. We have focused on the three-leptons   
signature which has a small Standard Model background, and 
allows a model-independent mass reconstruction
of the full sneutrino decay chain.\par  
A detailed study for an example MSSM point has shown that
the mass reconstruction analysis has an efficiency of
a few percent, and that a precise measurement of the masses of 
$\tilde\nu$, $\chionepm$, $\chione$
can be performed.  
Both the Standard Model background and the backgrounds
from other SUSY pair productions were studied in detail, and shown to be well 
below the expected signal for a value of the considered \rpv coupling 
$\lambda^{\prime}_{211}$ taken at the present low-energy limit. \par
The trilepton signal from sneutrino production 
was then studied as a function of the model parameters under different
model assumptions, and sensitivity over a significant part of the
parameter space was found. 
Within the MSSM, the production of
a 900~GeV sneutrino for $\lambda^{\prime}_{211}>0.05$, and of a 350~GeV sneutrino for
$\lambda^{\prime}_{211}>0.01$ can be observed 
in the first three years of LHC running. 
In the framework of the mSUGRA model, the region in the $m_0-m_{1/2}$ 
space accessible to the analysis was mapped as a function of the 
value of the $R_p$-violating coupling for representative values 
of $\tan\beta$. A significant part of the $m_0-m_{1/2}$ plane 
will be accessible for $\lambda^{\prime}_{211}>0.01$.\par
Although the detailed study was focused on the case of a single dominant 
$\lambda^{\prime}_{211}$ coupling,
we have found that the resonant sneutrino production analysis can bring
interesting sensitivities on all the \rpv couplings
of the type $\l'_{2jk}$,
compared to the low-energy constraints. 
The resonant sneutrino production 
should also allow to test most
of the $\l'_{1jk}$ coupling constants. \par 
In conclusion we have demonstrated that if minimal supersymmetry 
with R-parity violation is realised in nature, the three-leptons   signature
from sneutrino decay will be a privileged channel for the precision
measurement of sparticle masses and for studying the SUSY parameter space, 
over a broad spectrum of models.
Analyses based on the study of events including three leptons,  
were often advocated in the literature \cite{Nath}-\cite{Pai}  as a particularly 
sensitive way of attacking the search for SUSY at the LHC in the
standard R-conserving scenario. The higher lepton multiplicity and
the possibility to perform precise measurements of the model parameters
make this kind of analyses an even more attractive possibility 
in the case of R-parity violation with dominant $\lambda^{\prime}$
couplings.
\section*{Acknowledgements}
This work was initiated during a workshop held in Les Houches. We warmly thank
Patrick Aurenche and all of the organising team for the stimulating
program, the excellent atmosphere and the outstanding computing
facilities.
We are also deeply indebted to the HERWIG team for allowing us to use a
prerelease of the HERWIG 6.0  MonteCarlo.

\end{document}